\documentclass[aps,prl,twocolumn,superscriptaddress,nofootinbib]{revtex4}

\usepackage{graphicx}
\usepackage{epstopdf}
\usepackage{color}
\usepackage{bm}
\usepackage{todonotes}
\usepackage{amsmath}

\usepackage{setspace}
\let\oldmarginpar\marginpar
\renewcommand\marginpar[1]{\oldmarginpar[\vspace{-2cm}\singlespacing\raggedleft\scriptsize #1]{\vspace{-2 cm}\singlespacing\raggedright\scriptsize #1}}
\usepackage[normalem]{ulem}

\renewcommand{\Re}{\text{Re}}

\begin{document}
\bibliographystyle{apsrev}


\def\mb{\mathbf}

\newcommand{\ANGST}{\mbox{\normalfont\AA}}

\def\xcap{{\boldsymbol {\hat x}}}
\def\zcap{{\boldsymbol {\hat z}}}
\def\nn{{\nonumber}}
\def\n{{\bf n}}
\def\rad{{\bf r}}
\def\x{{\bf x}}
\def\E{{\bf E}}
\def\H{{\bf H}}
\def\B{{\bf B}}
\def\F{{\bf F}}
\def\p{{\bf p}}
\def\m{{\bf m}}
\def\P{{\bf P}}
\def\Q{{\bf Q}}
\def\K{{\bf K}}
\def\k{{\bf k}}
\def\u{{\bf u}}
\def\A{{\bf A}}
\def\s{\widehat{{\bf s}}}
\def\pp{\widehat{{\bf p}}}
\def\v{\mathbf{\Psi}}
\def\f{\mathbf{F}}
\def\g{{\tensor{\bf g}}}
\def\GG{{\tensor{\bm{ \mathbb{G}}}}}
\def\I{{\bf I}}
\def\al{{\tensor{\bm \alpha}}}
\def\ep{{\tensor{\bm \epsilon}}}
\def\G{{\stackrel{\leftrightarrow}{\bf{G}}}}  
\def\M{{\stackrel{\leftrightarrow}{\bf{M}}}}  
\def\GG{{\stackrel{\leftrightarrow}{\mathbb{G}}}}

\def\fermionlayer{2D-FL}
\def\fermionlayers{2D-FLs}
\def\thickness{d}

\newcommand\blfootnote[1]{%
  \begingroup
  \renewcommand\thefootnote{}\footnote{#1}%
  \addtocounter{footnote}{-1}%
  \endgroup
}


\title{Controlling surface charge and spin density oscillations by Dirac plasmon interaction in thin topological insulators}


\author{M. Ameen Poyli}
\affiliation{Centro de F\'{\i}sica de Materiales CFM - MPC, Centro Mixto CSIC-UPV/EHU, 20018 San Sebasti\'an-Donostia, Basque Country, Spain}
\affiliation{Donostia International Physics Center (DIPC), 20018 San Sebasti\'an-Donostia, Basque Country, Spain}
\affiliation{Department of Physics and Nanotechnology, SRM University, Kattankulathur, 603203, Tamil Nadu, India}

\author{M. Hrto$\mathrm{\check{n}}$}
\affiliation{Central European Institute of Technology, Brno University of Technology, Technick\'a 10, 616 00 Brno, Czech Republic}

\author{I. A. Nechaev}
\affiliation{Centro de F\'{\i}sica de Materiales CFM - MPC, Centro Mixto CSIC-UPV/EHU, 20018 San Sebasti\'an-Donostia, Basque Country, Spain}
\affiliation{Tomsk State University, Laboratory of Nanostructured Surfaces and Coatings, 634050, Tomsk, Russia}
\affiliation{Saint Petersburg State University, Laboratory of Electronic and Spin Structure of Nanosystems, 198504, Saint Petersburg, Russia}

\author{A. Y. Nikitin}
\affiliation{Donostia International Physics Center (DIPC), 20018 San Sebasti\'an-Donostia, Basque Country, Spain}
\affiliation{CIC nanoGUNE, 20018 Donostia-San Sebasti\'an, Spain}
\affiliation{IKERBASQUE, Basque Foundation for Science, 48013, Bilbao, Spain}

\author{P. M. Echenique}
\affiliation{Centro de F\'{\i}sica de Materiales CFM - MPC, Centro Mixto CSIC-UPV/EHU, 20018 San Sebasti\'an-Donostia, Basque Country, Spain}
\affiliation{Donostia International Physics Center (DIPC), 20018 San Sebasti\'an-Donostia, Basque Country, Spain}
\affiliation{Departamento de F\'{\i}sica de Materiales UPV/EHU, Facultad de Ciencias Qu\'{\i}micas, UPV/EHU, Apdo. 1072, 20080 San Sebasti\'an-Donostia, Basque Country, Spain}

\author{V. M. Silkin}
\affiliation{Donostia International Physics Center (DIPC), 20018 San Sebasti\'an-Donostia, Basque Country, Spain}
\affiliation{IKERBASQUE, Basque Foundation for Science, 48013, Bilbao, Spain}
\affiliation{Departamento de F\'{\i}sica de Materiales UPV/EHU, Facultad de Ciencias Qu\'{\i}micas, UPV/EHU, Apdo. 1072, 20080 San Sebasti\'an-Donostia, Basque Country, Spain}

\author{J. Aizpurua}
\affiliation{Centro de F\'{\i}sica de Materiales CFM - MPC, Centro Mixto CSIC-UPV/EHU, 20018 San Sebasti\'an-Donostia, Basque Country, Spain}
\affiliation{Donostia International Physics Center (DIPC), 20018 San Sebasti\'an-Donostia, Basque Country, Spain}

\author{R. Esteban}
\affiliation{Donostia International Physics Center (DIPC), 20018 San Sebasti\'an-Donostia, Basque Country, Spain}
\affiliation{IKERBASQUE, Basque Foundation for Science, 48013, Bilbao, Spain}

\email{ruben_esteban@ehu.eus} 



\begin{abstract}
Thin topological insulator (TI) films support optical and acoustic plasmonic modes characterized by effective net charge or net spin density, respectively. 
We combine many-body and electromagnetic calculations to study how these modes can be selectively excited  at films and nanodisks at infrared and THz frequencies.  
We first discuss the excitation of  propagating plasmons in a thin film by a point dipolar source. 
We emphasize how changing the distance between the dipolar source and the film allows to control the relative strength of the acoustic and optical plasmons and thus to excite net-spin or net-charge waves on demand.  
The acoustic and optical modes in a nanodisk structure can be efficiently tuned by changing the size of the disk or by applying electrostatic gating. 
Furthermore, these modes can be confined to regions of dimensions much smaller than the wavelength. 
The control of the excitation of acoustic and optical modes indicates that thin topological insulators are a promising system to manipulate the spin and charge properties of the plasmonic response, with potential applications in fast, compact and electrically-controlled spintronic devices.
\end{abstract}


\maketitle  

\section{Introduction}

\blfootnote{Copyright 2018 by the American Physical Society. Published in Phys. Rev. B, 97, 115420 (2018), https://journals.aps.org/prb/abstract/10.1103/\\ PhysRevB.97.115420}
Plasmonic resonances supported by metallic structures\cite{barnes2003,pitarke2006,pelton2008} allow to localize and manipulate light using very small structures, of dimensions well below the diffraction limit. Their resonant frequency is controlled by the overall geometry (shape, material...), and cannot be easily tuned after fabrication.  An alternative to plasmons in metals are Dirac plasmons \cite{DPlasmons} in  two-dimensional (2D) materials such as e.g. graphene\cite{Gr.THz.NatNano,nikitin16,fang13,yan12,mikhailov11}. These structures support plasmonic resonances that can be tuned  in the infrared and Terahertz frequencies by applying an external voltage (electrostatic gating), potentially reaching up to visible frequencies for systems just a few nanometers large\cite{G.Abajo.Gr.Review}. The excitation and tunability of 2D plasmons in graphene has been demonstrated in both near and far field optical measurements\cite{koppens_nl, jianing_nature, JablanPRB, Basov_gr,AuMono.G.Abajo}.

Tunable 2D plasmons are also present in three-dimensional topological insulator systems made of materials\cite{HaijunZhang} such as $\rm{Bi_2Se_3}$ and $\rm{Bi_2Te_3}$.  These topological materials are characterized by a bulk phase that is transparent in the infrared frequency range due to the aperture of a gap and by (quasi-) 2D fermion layers (\fermionlayers) at the interfaces with the surrounding medium\cite{SchutkyPRB, Hsieh-nature,Lai.TI.Plas,sim15}. As we do not consider 2D topological insulators presenting one-dimensional conduction channels\cite{hasan_Colloquium}, we simply call them topological insulators (TI) in the following.

The \fermionlayers\ of $\sim10\ \ANGST$ width are formed by surface-state electrons having a linear energy dispersion near the surface Brillouin zone center, and can support  surface plasmons at infrared and lower frequencies in a similar manner as graphene \cite{Kane_Mele_PRL, Wu_PRL,PRB_Lu, HaijunZhang, Hsieh-nature, Lai.TI.Plas}, as have been shown experimentally \cite{pietro_natnano13,uv-vis-topo,autore15,Deshko16,guozhi16}. Plasmons in TI can be tuned by electrostatic gating and present the additional advantage of carrying spin, which is locked with, and perpendicular to, the direction of propagation of the plasmon, and parallel to the surface at which they are excited. These excitations can thus be considered to be transverse spin waves\cite{RaghuPRL,Stauber14}.

Notably, the \fermionlayers~ at the two interfaces of a thin TI film can interact via Coulomb coupling. As a consequence, two hybridized modes appear, an acoustic mode characterized by an approximately linear dispersion relationship and an optical mode emerging at larger energies and showing a square-root dependence with wave number (for low frequencies)\cite{profumo12,Stauber12c}.  The optical (acoustic) mode is characterized by in- (out-of) phase charge waves induced at opposite surfaces of the thin TI film. Despite the similarity of these modes with those in spatially separated double-layer graphene\cite{hwang09}, the  real spin--momentum locking in TIs gives rise to a new phenomenon predicted for thin TI films---spin-charge separation. It implies that due to the opposite spin-momentum locking at the two surfaces of the film, the optical mode corresponds to an effective (pure) charge wave, while the acoustic mode demonstrates pure spin character\cite{StauberPRB,Stauber14,stauber17b}. Thus, thin films and other nanostructures made of a TI may serve as a nice playground both for studying the fundamental properties of collective excitations in the 2D Dirac fermions system and for engineering optoelectronic devices with tunable spin-dependent characteristics.

For practical applications in nanotechnology, it is important to understand how these modes can be excited and manipulated in an experiment, both for infinite films that support propagating plasmons and for localized plasmons in nanoparticles. 
The main objective of this paper is thus to study the selective excitation and control of  optical and acoustic modes in TI thin films and thin disks.
We use a combination of many-body calculations to obtain a non-local conductivity $\sigma$ characterizing each \fermionlayer\  (along with the plasmons dispersion of the systems) and classical electromagnetic calculations that take $\sigma$ as an input and provide the optical response under illumination by an external source. 
This approach also allows to study the importance of including non-locality when predicting the dispersion relationship of the systems.

After briefly introducing the dispersion relationship of a  thin film, we  analyze the excitation  of  propagating Dirac plasmons in such a system by localized point-like dipolar sources. We study the influence of the distance between the source and the film on  the  efficiency of the excitation of the optical and acoustic modes, which allows thus to manipulate the spin and charge properties of the excited surface wave.  The last section analyzes the excitation of  localized acoustic and optical plasmonic modes in thin disks, at frequencies that strongly depend on the size of the structures. We are particularly interested in the tunability of the system when changing the Fermi energy of the \fermionlayers\,, for example by applying an external voltage. When not stated otherwise, atomic units are used in the equations, i.e., $e = \hbar = m_e= 1$, where $e$ is the electronic charge, $\hbar$ is the reduced plank constant and $m_e$ is the mass of the electron.

\section{Systems and calculation methods}
\label{sec:topo_system}

\begin{figure}[tb]
\includegraphics[width=0.5 \textwidth]{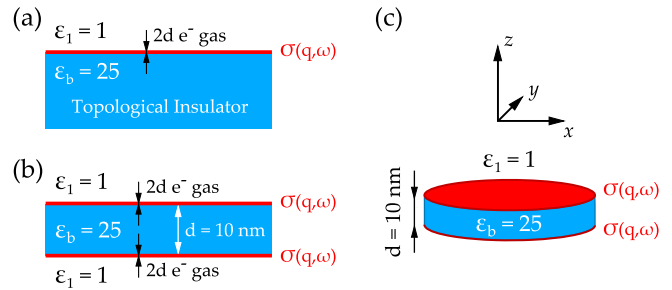}  \centering
\caption{Schematic of the TI structures considered.
(a) Semi-infinite TI substrate supporting a \fermionlayer\ at the interface with the surrounding vacuum, (b) an infinite TI film of thickness $\thickness=10$\,nm (suspended in vacuum) with two interacting \fermionlayers\ at the interfaces and (c) a finite TI thin disk  of thickness $\thickness$=10\,nm surrounded by vacuum, with a \fermionlayer\  at the upper interface and another one at the bottom. The 2D systems are characterized by an in-plane conductivity $\sigma(q,\omega)$.
We consider $\rm{Bi_2Se_3}$ as the topological insulator with relative dielectric permittivity of the bulk $\varepsilon_{\rm{b}}$=25. The center of the disk is at (0,0,0), and $z=0$ is the horizontal middle plane of the disk and the film in (b-c), in the coordinate system sketched in (c). We use a Fermi energy of the \fermionlayers\  $E_{\rm{F}}=250$\,meV, unless specified otherwise.  }
\label{schematic_1}
 \end{figure}

Fig. \ref{schematic_1} shows the schematics of the different TI systems, which are placed in vacuum (relative permittivity $\varepsilon_1$ = 1) and illuminated at angular frequency $\omega$ (vacuum wavelength $\lambda$ and vacuum wavenumber $k$ ).  We first briefly discuss the dispersion relationship of a semi-infinite substrate (Fig. \ref{schematic_1}(a)) and a thin film (Fig. \ref{schematic_1}(b)), focusing next on the near- and far-field response of the latter. 
Notably,  acoustic and optical propagating plasmonic modes can be excited in thin films by localized sources, with wavenumber $ q^{\rm{ac}}_{\rm{SP}}$ and $ q^{\rm{op}}_{\rm{SP}}$, respectively. We last discuss the excitation of acoustic and optical localized plasmonic resonances on finite thin disks (Fig. \ref{schematic_1}(c)).  For the near fields, it is  convenient to differentiate between  the total $E^{\rm{tot}}$ and the induced  $E^{\rm{ind}}$ amplitude of the electric fields, where only the former includes the fields from the illumination source itself.

The semi-infinite TI substrate in Fig. \ref{schematic_1}(a) supports a \fermionlayer\  at the interface with vacuum, which would be parallel to the (111) surface --the natural cleaving face of the considered TI. 
The TI thin-film  [Fig. \ref{schematic_1}(b)] and thin-disk  [Fig. \ref{schematic_1}(c)]  has a thickness $\thickness$=10\,nm and support two coupled \fermionlayers, one  at each flat interface (upper and bottom). The film extends infinitely in the lateral dimension, while the disks diameters considered are $D=40$\,nm and 300 nm. The flat surfaces of the film and the disk are perpendicular to the $z$ direction and parallel to the $x-y$ plane (see axes in the inset). $z=0$ corresponds to the central plane of the thin structures, with (0,0,0) the center of the disk.

We consider a simplified thin TI system surrounded by vacuum, with the objective of extracting general properties of the plasmon response under a given excitation. The response results from the electromagnetic coupling between the  \fermionlayers\ at the top and bottom interfaces. In this simple model,  we choose a typical value\cite{Dong_Science, Qi_APL, Glinka_APL, Sobota_PRL} for the relaxation time $\tau=500$\,fs and for the Fermi velocity of the electrons in the \fermionlayers\ of the TI $\rm{Bi_2Se_3}$, $v_{\rm{F}}=0.5\cdot 10^6$ m/s.\cite{xia2009observation} The bulk relative dielectric constant is taken as $\varepsilon_{\rm{b}}$=25, which is a reasonable approximation for  $\rm{Bi_2Se_3}$ in several regions of the THz spectrum\cite{IlyaPRB, PSSB.Stordeur,Kogar_2015arXiv,Bi2Se3_diel}. We thus ignore phononic contributions\cite{pietro_natnano13, autore15, autore15b, sim15} that may be significant in real experiments but that can complicate the phenomena of interest for this paper, i.e. the excitation of optical and acoustic modes. In the case of the disk calculations, an idealized model has also been chosen for the lateral sides, which are treated as not presenting any 2D conductivity. Further, we do not include corrections to the  electromagnetic constitutive  equations \cite{qi11,karch11}   (see Appendix B). We use a Fermi energy  $E_{\rm{F}}$ of the \fermionlayers\ equal to $E_{\rm{F}}$ = 250\,meV, unless other value is specified.

Our approach combines many-body and classical electrodynamic calculations. The many-body calculations described in Appendix A allow to calculate the dispersion relationship of the semi-infinite substrate and thin film, and to obtain the 2D in-plane conductivity $\sigma(q,\omega)$.  $\sigma$ depends on frequency and on the horizontal (parallel) wave-vector in the 2D layer, $q$. The $q$-dependence of $\sigma(q,\omega)$ captures non-local effects\cite{feibelman75,liebsch95,ruppin76,fuchs87} that modify the plasmonic response and that, in real-space, describe the dependence of the induced polarization at a given position on the electric fields  at the surrounding region. We assume that the two \fermionlayers~ of the film and disk  only interact electromagnetically (see Appendix A for details).

 The far- and near-field optical response under a given illumination is then calculated by solving the Maxwell's Equations, where each \fermionlayer\  is modeled by the 2D conductivity $\sigma(q,\omega)$.
 A dipole of strength 1 $e\cdot\,nm$ excites the  film and disk of $40$\,nm diameter, while a plane-wave incoming from the top illuminates the larger disks of $300$\,nm diameter. We describe here only the main ideas behind the calculations, which are discussed in more detail in Appendix B.  The effect of the conductivity of the \fermionlayers\ in this classical framework is to allow the accumulation of surface charge density ($\sigma_{\rm{c}}$) at the interfaces, which affects the boundary condition\cite{koppens_nl}.  The thin films are always excited by a point dipole (it is not possible to excite a plasmon in this case using far-field illumination), and we obtain the induced near-fields using a plane-wave decomposition method. By tracking the poles of the electromagnetic response (more exactly, the maximum of the transmission coefficient of the system) we also obtain a dispersion relationships that we compare with the results from the many-body model. This procedure allows to include the exact non-local $q$-dependence of the conductivity on the substrate and thin-film calculations. Furthermore, the local response can also be obtained by following the same procedure but always using $\sigma(q=0,\omega)$. Last, we obtain the response of the disks under plane-wave or dipolar excitation using full-wave calculations\cite{comsol}. Non-locality is introduced approximately into the calculations of the disks response by taking into account that each simulation is dominated by the response at a single $q$, corresponding to either the acoustic ($ q^{\rm{ac}}_{\rm{SP}}$) or the optical ($ q^{\rm{op}}_{\rm{SP}}$) mode.

\section{Excitation of propagating Dirac plasmons in infinite surfaces}

\begin{figure}[tb!]
\includegraphics[width=0.45 \textwidth]{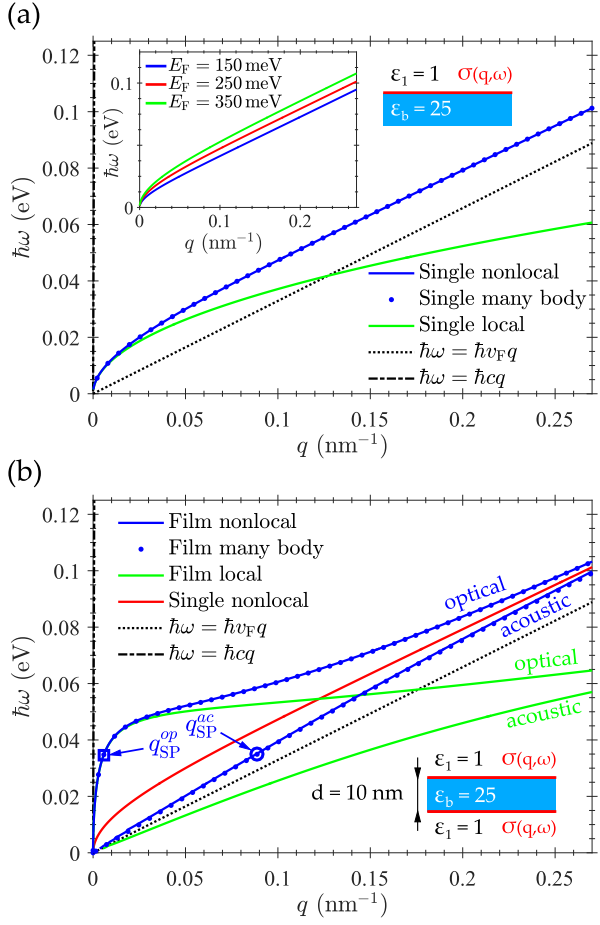}  \centering
\caption{Dispersion relationship of the propagating surface plasmons supported by (a) a semi-infinite TI substrate and (b) a thin film TI. (a) Results for the semi-infinite substrate, obtained from maximizing the strength of the classical electromagnetic response (i.e. the amplitude of the transmission coefficient) in $q$-space using local (green solid line) and non-local (blue solid line) conductivity, together with the non-local dispersion obtained directly from many-body calculations (blue dots). The inset shows the tunability of the non-local dispersion as a function of the Fermi energy.
(b)  Optical (high energy) and acoustic (low energy) branches of the dispersion relationships of Dirac plasmons in a thin film of thickness 10 nm. We show the results from maximizing the strength of the classical electromagnetic response in $q$-space using local (green solid lines) and non-local (blue solid lines)  conductivity, as well as the non-local dispersion obtained from many-body calculations (blue dots). For comparison, the non-local plasmon dispersion for a semi-infinite substrate in (a)  is also included in this panel (red solid line). Labels $q_{\rm{\rm{SP}}}^{\rm{op}}$ and $q_{\rm{SP}}^{\rm{ac}} $  mark the points of evaluation of the near-field in Fig. \ref{fig:nf_film} and charge and spin density in Fig. \ref{fig:charge_line_film} and Fig. \ref{fig:charge_density}. The relative dielectric function describing the bulk of the TI is $\varepsilon_{\rm{b}}$ = 25. The Fermi energy is $E_{\rm{F}} = 250$\,meV except in the inset of (a). For reference, the black short-dash line in (a) and (b) corresponds to $\omega = v_{\rm{F}} q$, with $v_{\rm{F}} = 0.5 \cdot 10^6$ m/s the Fermi velocity. The light cone is indicated by the dashed black line that is almost superimposed to the vertical axis.}
\label{disp}
\end{figure}

\subsection{Dispersion relationship}

We study in this section the excitation of propagating plasmons in TI systems that are infinite in the lateral direction (along the $x$ and $y$ directions indicated in the inset of Fig. \ref{schematic_1}). We are particularly interested on the thin films  depicted in  Fig. \ref{schematic_1}(b) but, for completeness, we first describe  the dispersion of plasmons in a TI semiinfinite layer (sketched in  Fig. \ref{schematic_1} (a)). The main panel in Fig. \ref{disp}(a)  and Fig. \ref{disp}(b) shows the dispersion relationship of plasmons in the substrate and thin film, respectively,  for $E_{\rm{F}}= 250$\,meV. We first obtain it by changing the angular frequency $\omega$ of an incoming (evanescent) plane wave and  tracking which parallel wavevector $q$ of this exciting field induces the strongest classical electromagnetic response (maximum in the transmission coefficient of the system). We consider both the full non-local [$\sigma(q,\omega)$, blue solid lines] and the approximated local [$\sigma(q=0,\omega)$, green solid lines] conductivity. The non-local results are in very good agreement with the dispersion obtained directly from the many-body calculations[blue dots].

The results for the semi-infinite substrate in Fig. \ref{disp}(a) reveal the presence of Dirac plasmons in the infrared-THz region of the spectrum that are characterized by a very large parallel wavenumber $q_{\rm{SP}}$ (low  wavelength $\lambda_{\rm{SP}}=2\pi/q_{\rm{SP}}$), compared to the corresponding value of the incident plane wave.  For example, a surface plasmon excited at $\lambda=35$\,$\mu$m ($\approx$ 0.0354\,eV) has a (non-local) plasmonic wavelength $\lambda_{\rm{SP}}\approx$ 95\,nm, showing a squeezing to $\lambda/$370. The field is thus evanescent and extremely confined in the vertical $z$ direction, i.e. the vertical wavenumber at vacuum ($k^1_z=\sqrt{(2\pi/\lambda)^2-q_{\rm{SP}}^2}$) and at the TI ($k^{\rm{TI}}_z=\sqrt{(2\pi/\lambda)^2\varepsilon_b-q_{\rm{SP}}^2}$) are imaginary and satisfy $\mid k^1_z \mid \approx  \mid k^{\rm{TI}}_z\mid \approx q_{\rm{SP}}  \gg 2\pi/\lambda $. For reference, the light cone is indicated by the almost-vertical black dashed line in Fig. \ref{disp}(a), which is characterized by wavenumber $2\pi/\lambda$.

The local and non-local dispersions are  similar for low energies.  In contrast, for sufficiently high energy (above $\approx 0.03$ eV), both curves in Fig. \ref{disp}(a) exhibit important quantitative and qualitative differences: in the case of the non-local approximation, the dispersion relationship becomes approximately linear, while for the local calculations the resonant energy is proportional to the  square root of $q_{\rm{SP}}$. As a result, the nonlocal slope   is $\approx 3$ times larger at the maximum $q_{\rm{SP}}$ shown, underlying the importance of using a non-local approach to accurately obtain the response for large energies. Further, the inset shows the strong dependence of the plasmon dispersion  on the Fermi energy $E_{\rm{F}}$ of the \fermionlayer, studied for the range $E_{\rm{F}} = 150$\,meV - 350\,meV in the non-local description. $E_{\rm{F}}$ can be experimentally controlled by an external voltage bias, so that the good tunability of the system constitutes a major advantage of 2D plasmons \cite{jianing_nature,  Wang_NLett_gating, Faxian-Nat.nano_gating}.

\begin{figure}[tb!]
\includegraphics[width=0.5 \textwidth]{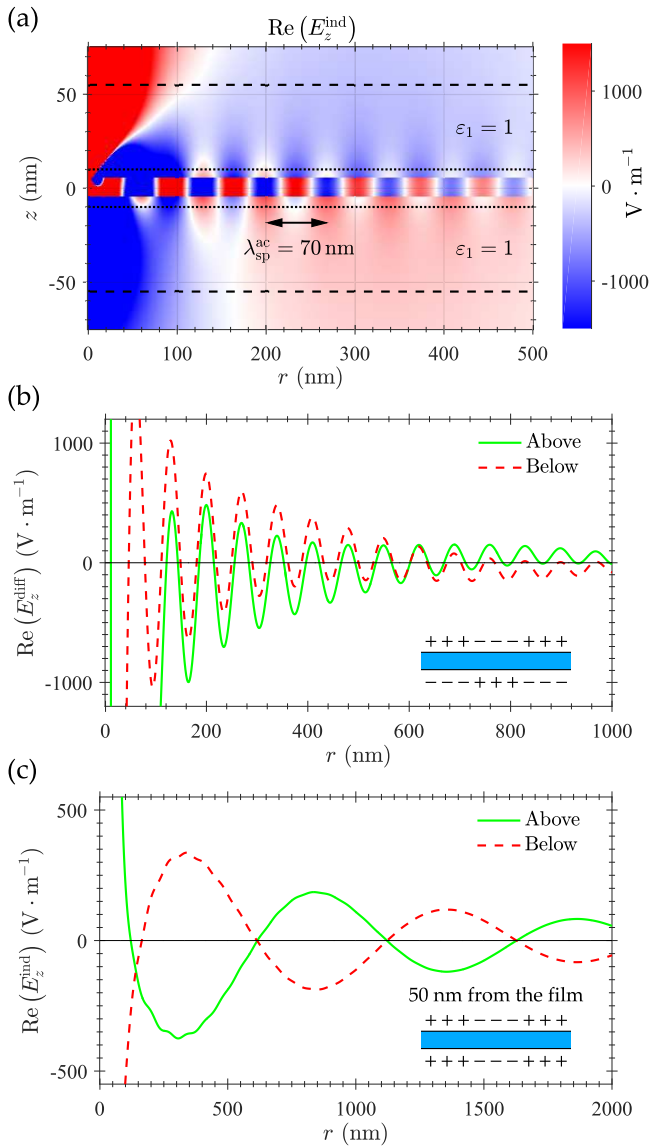}  \centering
\caption{ Fields  excited by a dipole illuminating a 10\,nm thin TI film. (a) Maps of the real part of the $z$ component of the electric field $E^{\rm{ind}}_z$ induced by a dipole emitting at $\lambda=$35 $\mu$m, situated 3\,nm above the film and oriented along the vertical $z$ direction. (b) Real part of the field in (a) at 3\,nm [short-dash lines in (a)] away from the film minus the corresponding value for 50\,nm distance [long-dash lines in (a)]. This subtracted field $E^{\rm{diff}}_z$ is calculated both for fields above [solid green line in (b)] and below [dashed red line in (b)] the thin film, following Eq. 1.
(c) Real part of the field calculated at 50\,nm distance from the upper (solid green line) and bottom (red dashed line) surfaces, along the dashed lines in (a). The insets in (b) and (c) indicate the charge distribution corresponding to the fields of the acoustic and optical modes, respectively
The Fermi energy is $E_{\rm{F}}$=250\,meV, the dipole strength is  $1e\cdot\,nm$ and the field is plotted in S.I. units (Vm$^{-1}$).  }
\label{fig:nf_film}
\end{figure}

Fig. \ref{disp}(b) shows  non-local (blue solid lines) plasmon dispersion for the $\thickness=10$\,nm thin TI film. The \fermionlayers\ at each interface interact with each other via Coulomb coupling and two hybridized plasmon modes emerge as two different branches in the dispersion, as studied in Ref. \onlinecite{StauberPRB}. The low and high energy branches of the dispersion fall at each side of the dispersion curve  for the semi-infinite substrate, (red line, corresponding to the non-local calculation). The mode associated with the lower energy branch exhibits an almost linear dependence between energy and wavenumber (linear dispersion relationship). It is usually called acoustic mode, and we will show that it is characterized by anti-symmetric charges at both interfaces (sketch in Fig. \ref{fig:nf_film} (b)). The acoustic plasmon is characterized by the largest plasmon parallel wavenumber  $q^{\rm{ac}}_{\rm{SP}}$ (smallest plasmon wavelength) and thus the strongest confinement  (largest $|k^1_z|$). On the other hand, the larger energy branch corresponds to the optical mode and present a symmetric charge distribution (sketch in Fig. \ref{fig:nf_film} (c)).
The  wavenumber $q^{\rm{op}}_{\rm{SP}}$ for the optical branch and energies smaller than $\approx 0.075$\,eV is significantly smaller than for the acoustic mode or for the plasmon of a single interface but remains much larger than that of a propagating plane wave in vacuum, so that the electromagnetic field of both acoustic and optical plasmonic modes are well confined to the TI film. For energies $\gtrsim 0.075$\,eV the acoustic and optical branches become similar to the dispersion relationship of a single interface.

As for the semi-infinite substrate,  the dispersion relationship at low energies change only weakly when comparing the results obtained using the local $\sigma(q=0,\omega)$ (green solid line) and non-local conductivity $\sigma(q,\omega)$ (blue solid line). Nonetheless, non-locality plays a  significant role in the high energy response. Looking first at the acoustic mode, the wavenumber $q^{\rm{ac}}_{\rm{SP}}$  is significantly larger for the local than for the more exact non-local approach.
The strong non-local effect can be explained by the conductivity  $\sigma(q^{\rm{ac}}_{\rm{SP}},\omega)$ at  very large  $q^{\rm{ac}}_{\rm{SP}}$ being significantly different from the local $\sigma(q=0,\omega)$ value. In the case of the optical branch,  the discrepancy between the local and non-local results becomes important only for energies larger than $\approx$ 0.05 eV, where the local calculation remains relatively flat as a function of $q$, with a much weaker slope than for the non-local conductivity.

\subsection{Fields and surface charges induced by a point dipole}

Due to their very large wavenumber $q_{\rm{SP}}\gg 2\pi/\lambda$,  a plane-wave cannot excite the  plasmonic modes in the TI film (or substrate) even when applying techniques often used to excite plasmons in flat  metallic surfaces,
such as exploiting the evanescent field induced by total internal reflection at the face of a prism (Otto configuration\cite{otto68}). We thus use a point-like dipolar source, which, as discussed in Appendix B, can be decomposed as an integral over plane-waves of different horizontal wavenumber $q$, with a significant contribution for a very large range of $q$ values  (very broad distribution in $q$-space). More exactly, we use  an electric dipole  oriented in the vertical $z$ direction and emitting at $\lambda$ = 35 $\mu$m.

We first consider a dipole positioned 3 nanometers above the TI thin film.
Fig. \ref{fig:nf_film}(a) shows the real part of the $z$ component of the electric field induced by the dipole, $\Re{(E}^{\rm{ind}}_z)$, in the vertical $x-z$ plane.
Due to the rotational symmetry of the dipolar illumination (with respect to the $z$ axis), the induced electromagnetic fields are rotationally symmetric as well. The oscillatory behavior of $\Re{(E}^{\rm{ind}}_z)$ in the radial $x$ direction  is typical of a  propagating surface wave.  We observe clear oscillations near(and inside) the film with periodicity $\approx 70$\,nm, very close to the plasmonic wavelength of the acoustic mode $\lambda_{\rm{SP}}^{\rm{ac}}$ obtained from the dispersion ($\lambda_{\rm{SP}}^{\rm{ac}}=2\pi/q_{\rm{SP}}^{\rm{ac}}$, with $q_{\rm{SP}}^{\rm{ac}}$ indicated by the blue open circle in Fig. \ref{disp}b). These fast oscillations clearly suggest the excitation of the acoustic plasmon.

The fast decay  in the vertical $z$ direction of these fast field oscillations indicate that they are strongly localized to a thin region near the TI film. The fields also decay along the direction of propagation ($x$ axis), but a large number of oscillations can be observed, revealing a large propagation distance (measured in units of the plasmonic wavelength). The decay of the plasmon in this direction is both due to the material losses associated with the  relaxation time of the TI and to the  $1/\sqrt{x}$ dependence  of the electric field of a lossless propagating plasmon  required by energy conservation (due to the cylindrical character of the plasmon wavefronts). A large propagation distance is desired for effective coupling and information transfer with distant systems.

As implied above, the $\approx 70$\,nm periodicity strongly supports the excitation of the acoustic plasmon\cite{StauberPRB},  characterized by fields oscillating in phase just above and below the TI (symmetric field distribution, corresponding to anti-symmetric charges as discussed below). According to the location of the maxima and minima of $\Re{(E}^{\rm{ind}}_z)$ in Fig. \ref{fig:nf_film}(a), however, one can appreciate that the phase of the fields is not exactly identical  at both sides of the film, and that the difference becomes larger in planes further away from the film (comparing the fields at $|z|$ to those at $-|z|$ for large $|z|$). Indeed, for planes sufficiently far away from the TI, the fields oscillate approximately with opposite phase above and below the film. This asymmetric field distribution corresponds to the expectation for optical modes\cite{StauberPRB}.

We attribute this behavior to the simultaneous excitation of both an optical and an acoustic propagating surface plasmon; the optical plasmon dominates the fields induced far away from the film, while the contribution of the acoustic plasmon become very important near the interfaces. More in detail, the excitation of the acoustic mode lead to the presence of the fast oscillations  near the film (set by $ q^{\rm{ac}}_{\rm{SP}}$) characterized by symmetric field with respect to the central plane. The large $q^{\rm{ac}}_{\rm{SP}}$ implies that the (evanescent) fields decay very fast in the vertical $z$ direction ($\mid k^1_z \mid  \approx  q^{\rm{ac}}_{\rm{SP}}$ ), so that it becomes difficult to appreciate these oscillations for $|z|$ larger than a few tens of nanometers.  In contrast, the  optical mode, characterized by  asymmetric field distribution and larger plasmonic wavelength, becomes dominant at sufficiently large $|z|$ due to its comparatively smaller $\mid k^1_z \mid \approx q^{\rm{op}}_{\rm{SP}} <  q^{\rm{ac}}_{\rm{SP}}$.

To confirm the simultaneous excitation of acoustic and optical modes with opposite field symmetries, we obtain $E_z$ at both sides of the film, at 3\,nm and 50\,nm away from it  (along the two short-dash and two long-dash lines in Fig. \ref{fig:nf_film}(a), respectively).
Since the  acoustic mode decays faster than the optical plasmon in the $z$ direction, we expect to be able to discriminate the fields associated with the former by subtracting the fields excited at the larger distance  to those at the smaller, i.e.,
\begin{eqnarray}
E^{\rm{diff}}_z=E^{\rm{ind}}_z(z=\frac{d}{2}+3\, \text{nm})-E^{\rm{ind}}_z(z=\frac{d}{2}+50\, \text{nm}) \nonumber \\ E^{\rm{diff}}_z=E^{\rm{ind}}_z(z=-\frac{d}{2}-3\, \text{nm})-E^{\rm{ind}}_z(z=-\frac{d}{2}-50\, \text{nm})
\end{eqnarray}
above and below the TI, respectively. Indeed, Fig. \ref{fig:nf_film}(b)  demonstrates that the fast oscillations of the subtracted signal $E^{\rm{diff}}_z$ above and below the film are in phase, as expected for the acoustic plasmon, and that the $\approx 70$\,nm periodicity is in good agreement with the value $2\pi/q^{\rm{ac}}_{\rm{SP}}$ obtained from the acoustic branch of the dispersion relationship (as marked by the blue open circle in Fig. \ref{disp}(b))

As we evaluate the fields further away from the film the relative weight of the optical mode increases. We plot in Fig. \ref{fig:nf_film}(c) the real part of $E^{\rm{ind}}_z$ at a distance of 50\,nm above and below the film [along the long-dash lines in Fig. \ref{fig:nf_film}(a)]. The fields have an opposite sign at each side of the film,  i.e. opposite orientation or $\pi$ phase difference, with a periodicity of the oscillations of $\approx$ 1000\,nm.
This anti-symmetry corresponds to symmetric charges (see below) and is typical of an optical mode in thin TI films \cite{StauberPRB}, and the periodicity matches well with the value $q^{\rm{op}}_{\rm{SP}}$ obtained from the corresponding branch of the dispersion relationship (marked by the blue square in Fig. \ref{disp}(b)). The fields in Fig. \ref{fig:nf_film} are thus consistent with the excitation of both an acoustic and optical mode, the former strongly contributing to the fields in the vicinity of the film and the latter predominating at positions further away.

\begin{figure}[tb!]
\includegraphics[width=0.5 \textwidth]{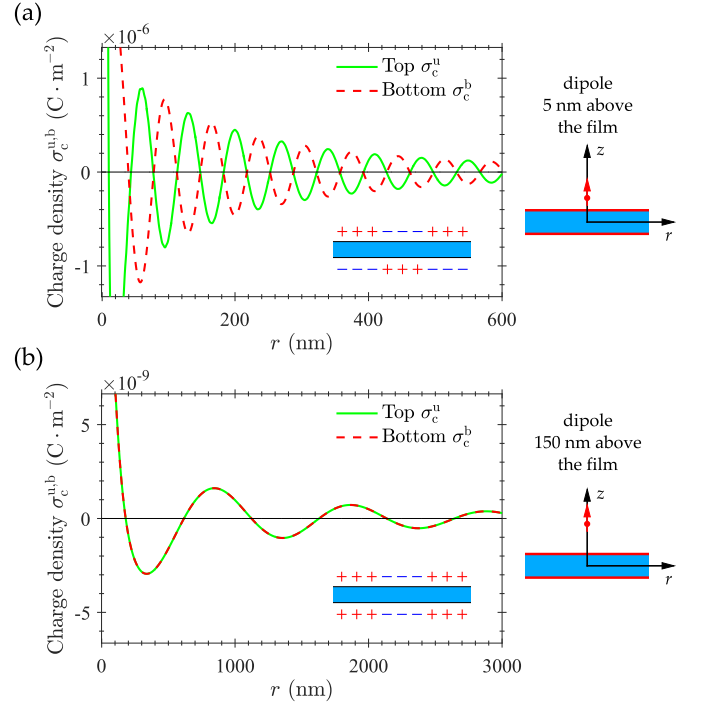}  \centering
\caption{Real part of the surface charge density in S.I units calculated along the $x$ axis  at the upper (green solid line) and bottom (red dashed line) interfaces of a TI thin-film of 10\,nm thickness.
A dipolar source radiating at wavelength $\lambda =35\,  \mu$m, placed at $x=y=0$ and oriented along the vertical $z$ direction is used to excite the surface plasmons in the film, for the Fermi energy $E_{\rm{F}}=250$\,meV . The charge densities, plotted for the dipole placed along the $z$ axis (a) 5\,nm and (b) 150\,nm above the film, show periodicities of $ \lambda_{\rm{SP}}\approx$  70\,nm and $ \lambda_{\rm{SP}}\approx$ 1000\,nm, respectively. The insets in (a) and (b) schematically show the corresponding charge distributions.}
\label{fig:charge_line_film}
\end{figure}

The field symmetries can be translated into charge symmetries through  the fields at both sides of each \fermionlayer . The surface charge density $\sigma_{\rm{c}}$ at the upper $\sigma^{\rm{u}}_{\rm{c}}$ and bottom $\sigma^{\rm{b}}_{\rm{c}}$ interfaces (in atomic units) are
\begin{eqnarray}
\sigma^{\rm{u}}_{\rm{c}} = \frac{\varepsilon_1}{4\pi} E^{\rm{tot}}_{1z}(z=\frac{\thickness}{2}) -\frac{\varepsilon_{\rm{b}}}{4\pi} E^{\rm{tot}}_{{\rm{TI}} z}(z=\frac{\thickness}{2}) \nonumber \\
\sigma^{\rm{b}}_{\rm{c}} = -\frac{\varepsilon_1}{4\pi} E^{\rm{tot}}_{1z}(z=-\frac{\thickness}{2}) +\frac{\varepsilon_{\rm{b}}}{4\pi} E^{\rm{tot}}_{{\rm{TI}} z}(z=-\frac{\thickness}{2})
\label{eq:D_sigma}
\end{eqnarray}
where ${E}^{\rm{tot}}_{1z}(z=\pm\frac{\thickness}{2})$ is the total field along $z$ on the vacuum side of the interface, and ${E}^{\rm{tot}}_{{\rm{TI}}z}(z=\pm\frac{\thickness}{2})$ the equivalent value inside the TI, both evaluated just at the film-vacuum interfaces. The sign difference between the equations of $\sigma^{\rm{u}}_{\rm{c}}$ and $\sigma^{\rm{b}}_{\rm{c}}$ is due to the dependence of the charge density on the direction of the surface normal. The anti-symmetric $E_z$ field exhibited by the optical modes thus corresponds to a charge distribution characterized by equal sign at both interfaces, which explains the comparatively larger energy of the optical branch in the dispersion relationship as a consequence of Coulomb repulsion between equal charges  at the upper and bottom \fermionlayers\ [inset of Fig. \ref{fig:nf_film}(c)]. In contrast, the acoustic modes present charges of opposite sign at the two TI-vacuum interfaces [inset of Fig. \ref{fig:nf_film}(b)].  As we will see in the next section, these charge symmetries are the opposite to those of the spin, which introduces a handle to control the net charge and net spin. Furthermore,  we also note that these charge symmetries are the opposite to those of the optical and acoustic plasmonic modes in a thin metallic film \cite{Nanohole.Plasmons}, where the difference between TIs and metal films appears because the latter exhibit a screening effect due to the 3-D electron gas (described by the negative value of the real part of the permittivity) that is not present in TIs.

\subsection{Controlled excitation of charge and spin waves by dipole-sample separation control}

Since the contribution to the fields induced by the dipole in Fig. \ref{fig:nf_film} from the acoustic and optical modes can be dominant near and far from the TI film, respectively, reciprocity\cite{carminati98} suggests that the acoustic mode would  be effectively excited by a dipolar source situated near the film, whereas the optical mode would be predominantly excited when the dipole is far away from the film.  With this in mind, Fig. \ref{fig:charge_line_film} shows the real part of the surface charge densities $\sigma_{\rm{c}}$ induced in the upper and bottom interfaces by a dipolar source radiating at $\lambda$ = 35\,$\mu$m situated at (a) 5\,nm and (b) 150\,nm away from the surface of the film.

When the dipole is at a distance of 5\,nm above the 10\,nm thin film [Fig. \ref{fig:charge_line_film}(a)] the induced charges at the upper and bottom surfaces has opposite sign and a spatial periodicity of $\approx 70$\,nm, indicating as expected the excitation of an acoustic mode. As an aside, Fig. \ref{fig:nf_film} showed that a dipole near the thin film also excite an optical mode that contributes to the near-fields outside the TI, but whose contribution to the charge distribution in Fig. \ref{fig:charge_line_film}(a) is difficult to appreciate; this contrast between field and charges is possible because  $\sigma_{\rm{c}}$ depends on the (displacement) fields outside and inside the film (Eq. \ref{eq:D_sigma}).

For a dipole placed 150\,nm above the surface, [Fig. \ref{fig:charge_line_film}(b)], the charge densities oscillate in phase at both interfaces and oscillate with the larger spatial periodicity of $\approx 1000$\,nm characteristic of the optical plasmon. The charges are thus indeed dominated by the acoustic mode for 5\,nm distance, and by the optical mode for 150\,nm. In consequence, it is possible to tune the relative strength of the two modes by varying the dipole-film distance. In this way, we can control not only the charge but also the spin properties, as we discuss in more detail below.


\begin{figure}[tb!]
\includegraphics[width=0.5 \textwidth]{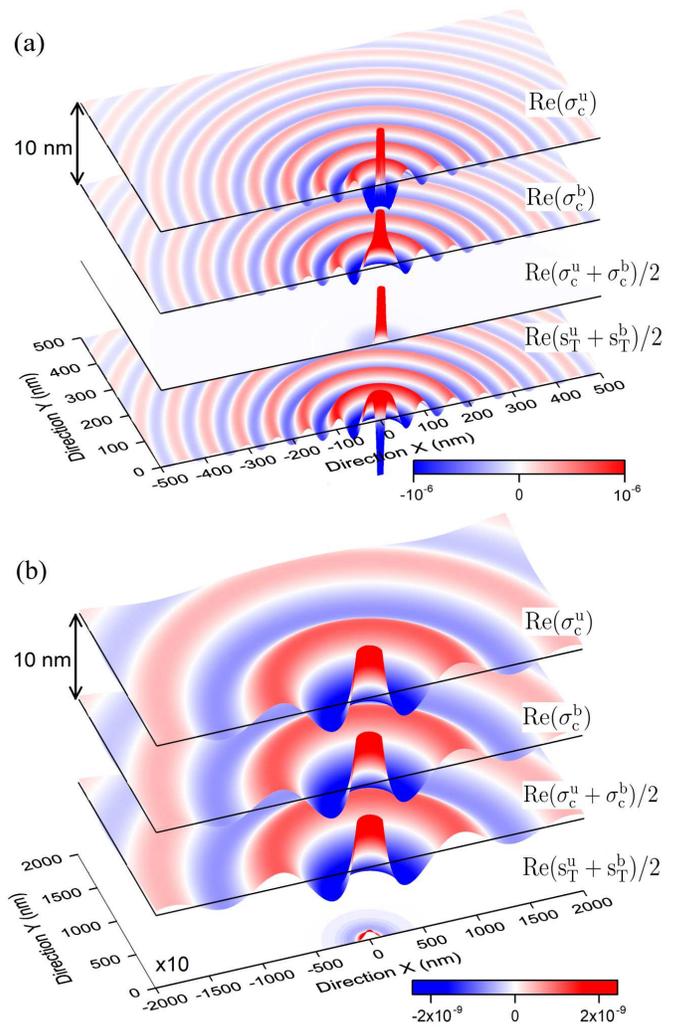}  \centering
\caption{Charge and spin waves in a $10$\,nm thin TI film for a dipole  placed at $x=y=0$ at two different distances above the film. The dipole  emits at $\lambda$ = 35 $\,\mu$m and is oriented along the vertical $z$ direction. (a)  Charge and spin density induced for the dipole positioned 5\,nm above the top surface of the film. The top two panels show the real part of the charge densities induced at the upper $\sigma_{\rm{c}}^{\rm{u}}$ (uppermost panel) and bottom $\sigma_{\rm{c}}^{\rm{b}}$ (panel below)  interfaces. The bottom two panels represent the real part of half the sum of the spin density  ($\Re(s_{\rm{T}}^{\rm{u}}$+$s_{\rm{T}}^{\rm{b}}))/2$, lowest panel) and charge density ($\Re(\sigma_{\rm{c}}^{\rm{u}}$ +$\sigma_{\rm{c}}^{\rm{b}})/2)$, panel above) at both interfaces, and thus indicate the behavior of (half) the effective charge ($\sigma_{\rm{c}}^{\rm{eff}}/2$) and spin ($s^{\rm{eff}}/2$) density oscillations, respectively.
(b) as in (a) but for 150\,nm distance between the dipole and the film. (a) and (b) correspond to the same systems showed in Fig. \ref{fig:charge_line_film} (a) and (b). Charge density is shown in S.I. units. The spin density is plotted in arbitrary units. The Fermi energy of the \fermionlayers\ is $E_{\rm{F}} = 250$\,meV. The values of $\Re(s_{\rm{T}}^{\rm{u}}$+$s_{\rm{T}}^{\rm{b}}))/2$ in (b) are multiplied by a factor 10.  }
\label{fig:charge_density}
\end{figure}

We analyze next the response in terms of net charge and net spin plasmonic waves.
The two top panels in Fig. \ref{fig:charge_density}(a) show the real part of the induced charge density oscillations at the upper ($\sigma_{\rm{c}}^{\rm{u}}$) and bottom ($\sigma_{\rm{c}}^{\rm{b}}$) \fermionlayer\, excited by the vertical dipole at 5\,nm above the film. As just discussed,  the surface charge density with $\approx 70$\,nm period is characterized by opposite phase at the upper and lower faces, corresponding to the acoustic plasmon wave. Thus, the total effective charge density $\sigma_{\rm{c}}^{\rm{eff}}$, defined as  $\sigma_{\rm{c}}^{\rm{eff}}=\sigma_{\rm{c}}^{\rm{u}}+\sigma_{\rm{c}}^{\rm{b}}$, is nearly zero.

Indeed, $\Re(\sigma_{\rm{c}}^{\rm{eff}})/2$, shown in the second panel from the bottom, is only comparable to either  $\Re(\sigma_{\rm{c}}^u)$ or $\Re(\sigma_{\rm{c}}^b)$ near the position of the dipole, where many of the evanescent components of the dipolar excitation contribute to the response, not only those associated with the excited plasmonic mode. On the other hand, a dipole at 150\,nm above the film, which excites predominantly the $\lambda^{\rm{op}}_{\rm{SP}}\approx 1000$\,nm optical mode, results in a strong total effective charge, as the charge densities at both surfaces of the film oscillate in phase (Fig. \ref{fig:charge_density}(b)).

The peculiarity of the TI film, compared, for instance to two graphene layers coupled electrostatically, is the surface-state spin  texture with well-defined helicity at the upper and bottom film surfaces. Thus, these TI films are potentially interesting for spintronic applications. The spin is parallel to the \fermionlayers, transverse to the direction of propagation of the excited plasmon and its direction depends on the surface normal.
More specifically, the transverse spin polarization density at the interfaces, of amplitude $s_{\rm{T}}$ couples with the charge density as given by $s_{\rm{T}}^{\rm{u,b}}(q,\omega)=\mp\frac{\omega}{v_{\rm{F}}q}\sigma^{\rm{u,b}}_c(q,\omega)$ for the upper ($s_{\rm{T}}^{\rm{u}}$, minus sign)  and bottom ($s_{\rm{T}}^{\rm{b}}$, plus  sign) surfaces\cite{RaghuPRL,StauberPRB}. Thus, for the in-phase charge oscillations characteristic of the optical mode the transverse spin polarization have opposite direction at the two 2D-FLs. In contrast, the acoustic  resonance is characterized by the same spin-polarization direction at both interfaces.

 To provide a more precise characterization of the behavior of the effective spin, we calculate the spin density amplitude as a function of position $\mathbf{r}$ by performing the Fourier transform of $s_{\rm{T}}^{\rm{u,b}}(q,\omega)$. We then  plot $s^{\rm{eff}}/2= (s_{\rm{T}}^{\rm{u}} + s_{\rm{T}}^{\rm{b}})/2$ in the bottom panels of  Fig. \ref{fig:charge_density}(a,b), which confirms that $s^{\rm{eff}}$ is large when the net charge density $\sigma_{\rm{c}}^{\rm{eff}}$ is small and vice versa. In short, the acoustic and optical modes can be seen as spin-like or charge-like density waves, respectively, whose relative strength is controlled by the position of the dipole; specifically,  as the dipole is displaced from very short to very large distance from the TI film, we transition from exciting spin collective excitations to inducing charge density  waves.

\section{Localized surface plasmons in $\bf{\rm{Bi_2Se_3}}$ nanodisks}
\label{sec:topo_disk}

 We move next from studying propagating plasmons in infinite TI surfaces to the excitation of localized Dirac plasmons in finite TI structures, such as the nanodisks of diameter $D$ depicted in Fig. \ref{schematic_1}(c). These disks can act as optical nanoantennas enhancing and confining the field, leading for example to strong interactions with nearby objects or molecules\cite{hanarp03,esteban08,lassiter10,esteban10}. As we illustrate below, the rupture of translational symmetry also allows the excitation of (optical) plasmonic modes by an incoming plane wave.

 Similarly to what has been studied for metallic structures\cite{bozhevolnyi07,novotny07}, plasmonic resonances in disks can be understood as Fabry-P\'erot-like cavity modes, for which the reflection of propagating plasmon at the edges leads to localized resonances when constructive interference occurs.
The disks support many different resonances\cite{kuttge10,fang13,volkov88}, and, for example, the difference between edge and sheet plasmons in graphene and metallic disks have been recently studied\cite{wang11,schmidt14,nikitin16}. For simplicity, we focus on  configurations where only a few of the possible modes are excited. More in detail,  we  choose illumination sources that induce  modes where the charge densities induced at the \fermionlayers\  are antisymmetric  with respect to the vertical $x=0$ plane (not to be confused with the symmetry with respect to the middle of the film discussed in the previous section) and present a cosine-like evolution in the polar direction (in the $xy$ plane). In a simple model treating the plasmon as a plane wave-like surface excitation of well defined wavevector $q_{\rm{SP}}$ and assuming zero reflection phase $\phi_r$ of the plasmon at the edges of the disk of diameter $D$\cite{gordon06,barnard08}, these resonances emerge when $q_{\rm{SP}}$ verifies
\begin{equation}
q_{\rm{SP}}=\frac{\pi n}{D}
\label{eq:cavitymodel1}
\end{equation}
with $n\geq1$ an odd integer (even values of $n$ describe rotationally symmetric resonances not excited by our illumination scheme). Eq. \ref{eq:cavitymodel1} corresponds to the condition of coherent interference that occurs when the plasmon accumulate a $2\pi n$ difference in a full round trip. Refs. [\onlinecite{filter12,tzerkezis15}]  use a related model in terms of Bessel Functions that gives a somewhat different $q_{\rm{SP}}$ but does not change the general conclusions below.  Once $q_{\rm{SP}}$ is known from Eq. \ref{eq:cavitymodel1}, the resonant energy is obtained via the dispersion relationship. Notice that considering a non-zero reflection phase $\phi_r$ would modify to some extent the position of the resonances with respect to the prediction by this simplified equation.

In the following sections, we consider the  thin disks sketched in Fig. \ref{schematic_1}(c) with thickness $d = 10$ nm and diameter  $D = 40$\,nm or $D = 300$\,nm.

\begin{figure}[tb!] 
\includegraphics[width=0.5 \textwidth]{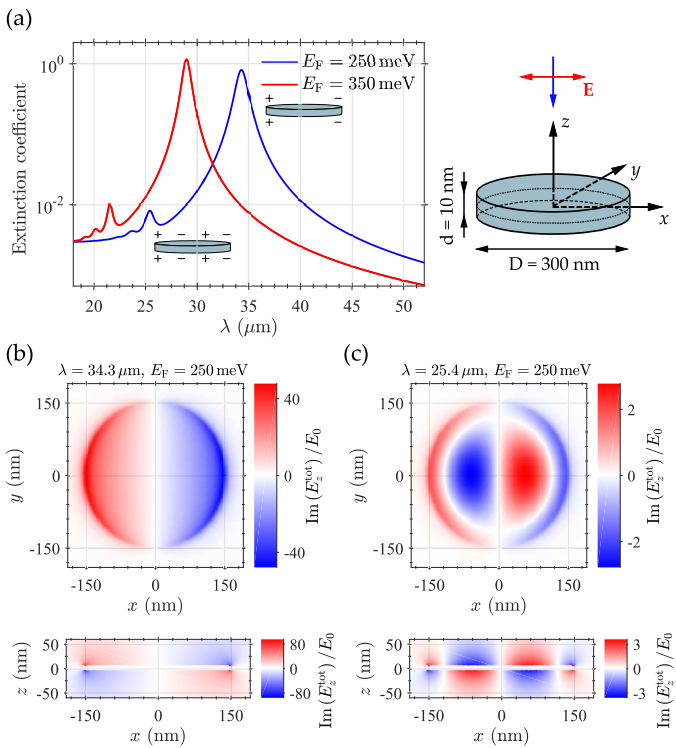}  \centering
\caption{
Optical response of a thin TI disk of diameter 300\,nm and thickness $\thickness=10$\,nm that supports localized optical plasmons. The excitation is a plane wave polarized along the $x$ direction and incident from the top ($z$ axis), as shown in the schematic (top right). The coordinate axis indicates the different directions, and the center of the disk is situated at  $x$ = 0, $y$ = 0, $z$ = 0.
(a) Extinction coefficient of the disk, for  the Fermi energy $E_{\rm{F}} = 250$\,meV (blue solid line) and $E_{\rm{F}} = 350$\,meV (red solid line) of the \fermionlayers. (b) Maps of the imaginary part of the $z$ component of the total electric field normalized to the amplitude of the incoming wave, Im$(E_z^{\rm{tot}})/E_0$ , at the prominent lowest-energy spectral peak at $\lambda \approx 34.3$\,$\mu$m for $E_{\rm{F}} = 250$\,meV.  In the top panel, the fields correspond to the horizontal plane $z=10$\,nm, parallel to the disk and $5$\,nm over the top interface. The fields in the bottom panel are obtained in the $y$ = 0 vertical plane passing through the center of the disk and parallel to the incident electric field. (c) as in (b), but for the second lowest-energy peak at $\lambda \approx 25.4$\,$\mu$m in the $E_{\rm{F}} = 250$\,meV spectrum. }
\label{fig:ff_nf_disk600}
\end{figure}

\subsection{ Excitation of localized optical modes with plane wave illumination}
Fig. \ref{fig:ff_nf_disk600}(a) shows the extinction coefficient  (extinction cross section normalized by the disk top surface area) of the disk of $300$\,nm diameter,
when illuminated by a plane wave of amplitude $E_0$ incident from the top with linear polarization along $x$, as schematically shown in the figure. This polarization imposes anti-symmetry of the induced charges  in the direction of polarization, i.e. with respect to $x=0$.
Due to the symmetry of the structure with respect to the horizontal plane $z=0$  and the negligible phase shift associated with the propagation of the plane-wave across the small thickness of the disk, the charge densities at opposite points at the upper and bottom flat TI-vacuum interfaces must be equal, which corresponds to the excitation of optical modes. We discuss in Appendix B how  we introduce non-locality into the full-wave calculations, under the assumption that only the optical mode is excited, by using $\sigma(q^{\rm{op}}_{\rm{SP}})$, the conductivity  evaluated at $q^{\rm{op}}_{\rm{SP}}$.

The extinction spectra (Fig. \ref{fig:ff_nf_disk600}a) show clear peaks associated to the localized plasmon resonances.
For $E_{\rm{F}}=250$\,meV (blue solid line), the two lowest energy peaks are found at $\lambda = 34.3$\,$\mu$m and $\lambda = 25.4$\,$\mu$m. For comparison, the simple Eq. \ref{eq:cavitymodel1} using $n=1,3$ predicts resonances at $\lambda = 30.7$\,$\mu$m and $\lambda = 25.5$\,$\mu$m,  in reasonable agreement with the simulated results. The broadness of a peak at frequency $\omega_{\rm{PL}}$ can be quantified with the quality factor $Q=\omega_{\rm{PL}}/\Delta\omega$, where $\Delta\omega$ is the frequency difference corresponding to the full-width half maximum of the peak. The quality factor of the $\lambda = 34.3$\,$\mu$m peak is $Q=26.9$, comparable to the values found in  plasmonic systems\cite{wang06c}. It is  in good agreement with the result of the quasistatic equation\cite{wang06c} $Q=\omega^2  \frac{d \left[\text{Im}(\sigma(q^{\rm{op}}_{\rm{SP}}))/\omega\right]}{d\omega}/(2\text{Re}(\sigma(q^{\rm{op}}_{\rm{SP}})))\approx 27$ (evaluated at the resonant frequency), which only depends on the real and imaginary part of the conductivity, not on the exact geometry.

The extinction peaks shift significantly for moderate changes of the Fermi energy $E_{\rm{F}}$ (which modifies $\sigma(q^{\rm{op}}_{\rm{SP}},\omega$)). The lowest energy peak, for example, shifts from $\lambda = 34.3$\,$\mu$m to
$\lambda = 29$\,$\mu$m when $E_{\rm{F}}$ changes from $250$\,meV (blue solid line) to  $350$\,meV (red solid line). The energy shift is $\approx 4$ times the full with at half maximum $\Delta\omega$ of the  $E_{\rm{F}}=250$\,meV peak, illustrating the high tunability of plasmonic resonances in TI disks.

In order to confirm the character of the spectral peaks, we calculate the electric field distribution around the disk at the position of the two lowest energy peaks for the $E_{\rm{F}} = 250$\,meV spectrum (blue spectrum in Fig. \ref{fig:ff_nf_disk600}a). Fig. \ref{fig:ff_nf_disk600}(b) shows the imaginary part of the $z$ component of the amplitude of the total electric field, normalized by the amplitude of the incident plane wave $E_0$, Im($E_z^{\rm{tot}})/E_0$, for the $\lambda = 34.3$\,$\mu$m resonance. Fig. \ref{fig:ff_nf_disk600}(c) presents Im($E_z^{\rm{tot}})/E_0$ for the resonance excited at $\lambda = 25.4$\,$\mu$m. The fields are calculated in the horizontal plane $5$\,nm above the disk (corresponding to $z=10$\,nm, top panels) and in the  $y$=0 plane of incidence of the incoming illumination going through the center of the disk  (bottom panels). We can see that the fields are symmetric with respect to the $z=0$ horizontal plane.  In consequence (see Eq. \ref{eq:D_sigma}), the induced charges are anti-symmetric with respect to the same plane, i.e. same sign at opposing interfaces [insets in Fig. \ref{fig:ff_nf_disk600}(a)],
as anticipated for the excitation of an optical mode. Furthermore, both the charges and fields are  antisymmetric  along $x$, with a cos-like symmetry in the polar direction. Last, the field distribution directly over the disk present 1 and 3 nodes along the $x$ axis for the peaks at $\lambda = 34.3$\,$\mu$m and  $\lambda = 25.4$\,$\mu$m, respectively. Therefore, the lowest energy plasmon at $\lambda = 34.3$\,$\mu$m corresponds to a dipolar optical mode, and the resonance at $\lambda = 25.4$\,$\mu$m is a higher order optical mode. Our results thus confirm the excitation of tunable optical plasmonic  resonances in TI disks by an incident plane wave.

\begin{figure}[tb!] 
\includegraphics[width=0.4 \textwidth]{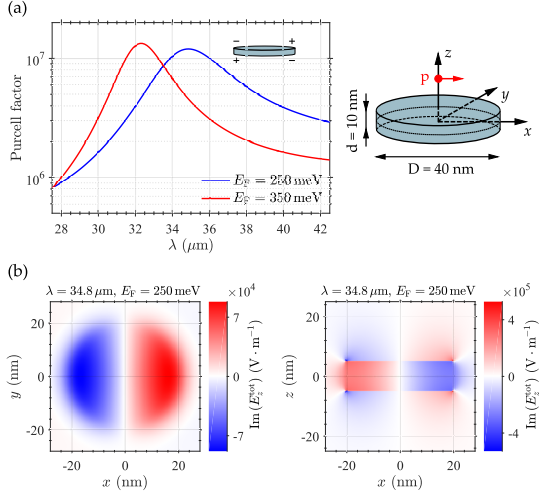}  \centering
  \centering
\caption{Optical response of a thin TI disk of diameter 40\,nm and thickness $\thickness=10$\,nm that supports localized acoustic plasmons. The excitation is an electric dipole situated $5$\,nm above the center of the top surface of the disk ($z=10$\,nm) and oriented  along the horizontal $x$  direction, as shown in the schematic (top right). The coordinate axis indicate the different directions and the center of the disk is situated at $x$=0, $y$=0, $z$=0.
(a) Purcell factor calculated for the Fermi energy $E_{\rm{F}}=250$\,meV (blue solid lines) and $E_{\rm{F}}=350$\,meV (red solid lines) of the \fermionlayers. (b)
 Maps of the imaginary part of the $z$ component of the total electric field Im$(E_z^{\rm{tot}})$, at the  lowest-energy spectral peak at $\lambda = 34.8$\,$\mu$m for $E_{\rm{F}}=250$\,meV.  In the left panel, the fields correspond to the horizontal plane $z=10$\,nm, parallel to the disk and $5$\,nm over the upper interface. The fields in the right panel are obtained in the $y$=0 vertical plane, passing through the center of the disk and parallel to the dipole orientation. The strength of the dipole, which does not affect the Purcell Factor in (a), is set to $1 e\cdot \mathrm{nm}$ in (b).}
\label{fig:ff_nf_disk50}
\end{figure}

\subsection{Excitation of a localized acoustic mode by a point dipole}

We consider next a disk of almost ten times smaller diameter $D=40$\,nm and same $\thickness=10$\,nm thickness. Using this smaller diameter allows to obtain the lowest-energy acoustic plasmon at a similar frequency as the corresponding optical mode for the larger disk.
 We are interested in the excitation of acoustic modes, so that we use the conductivity $\sigma(q^{\rm{ac}}_{\rm{SP}},\omega)$ in the full wave calculations (see Appendix B). As we discuss below, we verify that the modes emerging in the simulations are indeed acoustic. Furthermore, we have also verified (not shown) that using $\sigma(q^{\rm{op}}_{\rm{SP}},\omega)$ in the calculations to correctly describe the optical modes  would not introduce any additional peak in the wavelength range studied $\lambda> 27.5$\,$\mu$m (the lowest-energy optical peaks appear at $\lambda=23.7$\,$\mu$m and  $20.2$\,$\mu$m for the two values of the Fermi energy considered). Thus, using $\sigma(q^{\rm{ac}}_{\rm{SP}},\omega)$ instead of the non-local $\sigma(q,\omega)$ should be adequate for our study of the excitation of acoustic resonances.

As previously discussed, a plane wave exciting the thin disks cannot efficiently excite the anti-symmetric charge distribution (opposite sign at the top and bottom interface) characteristic of acoustic modes. Thus, we consider excitation by a dipole 5\,nm above the center of the top surface of the disk, oriented parallel to this surface (along $x$). The spectra in Fig. \ref{fig:ff_nf_disk50}(a) shows the Purcell Factor, for the same Fermi energies as those considered in Fig. \ref{fig:ff_nf_disk600}.
The Purcell Factor $P_{\rm{F}}$ is defined here as the power emitted by the dipole in the presence of the disk normalized to the corresponding value for the dipole in vacuum, can be calculated from the induced fields as discussed in Ref. \onlinecite{novotnyBook} and does not depend on the dipole strength. A clear peak emerge in the spectra, broader than for the optical modes in Fig. \ref{fig:ff_nf_disk600} because the large $q^{\rm{ac}}_{\rm{SP}}$ implies proximity to the electron-hole continuum where losses increase.

For $E_{\rm{F}}=250$\,meV (blue solid line) the dominant  peak appears at $\lambda=34.8$\,$\mu$m  and has a quality factor $Q=\omega_{\rm{PL}}/\Delta\omega\approx 7.0$.
The resonant wavelength is again relatively close to the prediction $\lambda=39.7$\,$\mu$m given by  Eq. \ref{eq:cavitymodel1}. The tunability of the excited resonance as $E_{\rm{F}}$ changes from $250$\,meV (blue solid line) to $350$\,meV (red solid line) is  smaller than for the optical modes, but remains comparable to the FWHM of the peaks. The very large Purcell Factors (up to more than $10^7$) indicate that the plasmonic disk couple very efficiently with the dipole; they are a consequence of the strong localization of the fields in the horizontal and vertical direction (extremely low mode volume), as set by the lateral dimensions of the disk and by $k^1_z$, respectively.  Applying the simple equation for the Purcell Factor $P_{\rm{F}}=3/(4\pi^2)\lambda^3 Q/V_{\rm{eff}}$, we estimate an effective volume $V_{\rm{eff}}\approx (120\,\text{nm})^3$. Notably, we have assumed in Fig. \ref{fig:ff_nf_disk50} that the system remains on the weak coupling regime, where the effect of the nanodisk on a dipolar source is well described by the Purcell Factor. However,  such  small $V_{\rm{eff}}$ indicate that it could be possible  to reach the strong coupling regime even for sources with relatively small dipolar strength\cite{torma2014}

Fig. \ref{fig:ff_nf_disk50}(b) shows Im($E^{\rm{tot}}_z)$ for the lowest-energy peak  in the $E_{\rm{F}}=250$\,meV spectra.
The left and right panels represents, respectively, the fields in the horizontal plane $z=10$\,nm (5nm above the top interface) and the vertical plane $y=0$ panel (same planes as in Fig. \ref{fig:ff_nf_disk600}). The strength of the dipole is $1e\cdot$nm. The main difference between the results obtained for the larger disks under plane wave illumination (Fig. \ref{fig:ff_nf_disk600}) and for the smaller disk excited by a dipole (Fig. \ref{fig:ff_nf_disk50}(b)) is that the fields induced in the latter case are symmetric along the vertical direction, i.e. with respect to the central $z$=0 horizontal plane. According to Eq. \ref{eq:D_sigma}, the symmetric fields correspond to anti-symmetric charge distribution (sketched in the inset of Fig. \ref{fig:ff_nf_disk50}(a)) confirming the excitation of acoustic modes.
Furthermore, the charges (as also $E^{\rm{ind}}_z$) are antisymmetric with respect to $x=0$, with the charges excited at the upper interface forming a dipole-like pattern, which reminds the results for the large disks  in Fig. \ref{fig:ff_nf_disk600}.
However, as the sketch in Fig. \ref{fig:ff_nf_disk50} indicates, the charges exited at the bottom interface correspond to a dipole-like pattern of opposite sign than the one at the top. Thus, the overall distribution of the induced charges is quadrupolar. A quadrupole is weakly radiative (non-radiative in the ideal case) which is a different way of arguing why plane-waves do not couple efficiently with this acoustic mode.

We have thus shown that,  by choosing adequately the size and illumination of the disks, we can excite modes that have either acoustic or optical character. In a similar manner as shown in Fig. \ref{fig:charge_density} for the propagating waves, these localized modes would be dominated, respectively, by the value of the effective total spin density $s_{\rm{T}}^{\rm{eff}}$ or total charge density $\sigma^{\rm{eff}}_{\rm{c}}$. Modifying $E_{\rm{F}}$ allows to tune the resonant energy to a desired value.

\section{Summary and Discussion}
We have combined many-body theory and classical electrodynamic calculations to analyze in detail the excitation of propagating and localized plasmons supported by thin TI structures. To focus on the pure electrodynamic interactions between the plasmons at the upper and bottom interfaces, we consider a simple case where the bulk of the topological insulator is perfectly transparent. Thus, we do not include the contribution from bulk plasmons \cite{politano15} or from  phonons\cite{pietro_natnano13,autore15,autore15b,sim15}. These effects can increase the losses and modify the dispersion, possibly introducing coupling between surface and bulk modes that can lead to phenomena such as Fano resonances or strong coupling.  In the case of thin disks, edge effects\cite{G.Abajo.Gr.Review, G.Abajo.ACSNano} and the complex atomic structure of the TI side surfaces\cite{virk2016dirac}, where a \fermionlayer\  may form, could also shift and broaden the resonances. We also do not consider the depletion layer\cite{stauber17b}. On the other hand,  we have introduced the dependence of the 2D-conductivity on the parallel wave-vector $q$ in order to show how the strong non-locality of the response affects the optical response.

Thin TI films support optical and acoustic propagating plasmon modes. Acoustic plasmons are associated with net effective spin and vanishing total charge densities, while the optical mode is characterized by net charge oscillations with insignificant effective spin density.  We have studied how to excite this system, showing that the experimental condition strongly affects  the contribution of these two types of modes to the total signal.  Specifically, by changing the distance between a point-like dipolar source and the TI, we are able to strongly modify the relative weight of the acoustic or the optical plasmons, and thus to control the spin and charge character of the propagating surface wave. The dipolar excitation we considered  could come from a 2-level transition in single photon emitters, but finding appropriate transitions at these frequencies can be challenging. A tip, a small particle or any edge might serve as alternative localized sources\cite{novotny97,ocelic05}. It is also worth noting here that the approach of coupling double-2D-FL structures to control two different properties of the excitated surface wave can be extended to other situations. For example, Ref.~\onlinecite{Stauber_PRL_2018} discusses the separation of purely magnetic and charge plasmons in two twisted bilayer graphene structures.

Further, we show that, by controlling the geometry and illumination, localized acoustic or optical plasmons can be selectively excited in thin TI disks at well-defined wavelengths. These modes are confined to very small volumes, as desired for large coupling strengths and quantum applications \cite{bellessa04,trugler08,schmidt15}.  We have focused on separately exciting either acoustic or optical modes in the thin disks, but we have also verified that it is possible to excite both types of modes in the same disk at a similar frequency, (for example by exciting $D = 95$\,nm disks in the $\lambda = 28$\,$\mu$m range with a dipolar excitation). Additionally, we have assumed that the Fermi energy $E_{\rm{F}}$ is the same for the \fermionlayers\ of the thin structures, but it may be also of interest to explore the case where the applied voltage leads to two different $E_{\rm{F}}$.

 Notably, the modes supported by the thin disks are relatively narrow spectrally, and can be easily tuned by changing the geometry or the external voltage. They thus offer an attractive path to control the nett spin and charge excited in nanostructures, by selectively exciting acoustic or plasmonic modes, or both simultaneously. The possible excitation of a well-defined net spin is a key difference with respect to other similar systems, such as bilayer graphene.

In summary, we have demonstrated the large flexibility offered by thin TIs to control spin and charge properties of plasmonic resonances,
making them an attractive possibility to engineer very compact and fast \cite{autore15} optoelectronic and spintronic devices.


\subsection{ Acknowledgments}

M. A. P, J. A and R. E acknowledge funding from project FIS2016-80174-P of the Ministry of Economy, Industry and Competitiveness MINEICO and from  NIST grant 70NANB15H32 of the Department of Commerce of the US. M. H. acknowledges financial help from Grant Agency of the Czech Republic (grant No. 15-21581S), Technology Agency of the Czech Republic (grant No. TE01020233) and MEYS CR (project No. LQ1601 CEITEC 2020).  I.A.N. and V. M. S. acknowledge support from the Spanish Ministry of Economy, Industry and Competitiveness MINEICO (Project No. FIS2016-76617-P) and from Saint Petersburg State University (Grant No. 15.61.202.2015). A.Y.N acknowledges the financial support from the Spanish Ministry of Economy, Industry and Competitiveness (national projects MAT2014-53432-C5-4-R and MAT2017-88358-C3-3-R).
R. E and V. M. S   acknowledge project PI2017-30 of the Departamento de Educaci\'on, Pol\'itica Ling\"u\'istica y Cultura of the Basque goverment.

\section {Appendix A: Many-body calculations}
\label{appendixmethod}

We describe first the many-body calculation of the conductivity $\sigma(q,\omega)$ required by the classical calculations (Appendix B) and obtained
for a single  \fermionlayer\ between the TI and vacuum. We express the conductivity in terms of the non-interacting response function $\chi^0_{\tau}(q,\omega)$ of a single \fermionlayer\ as
\begin{equation}
\sigma(q,\omega) = i\omega \; \frac{\chi^0_{\tau}(q,\omega)}{q^2}.
\label{eps_sigma}
\end{equation}
In the relaxation-time approximation, the non-interacting response function reads\cite{Mermin.PRB}
\begin{equation}
\chi^0_{\tau}(q,\omega) =  \frac{ \left(1+i/\omega \tau \right)\chi^0(q,\omega+i/\tau)}{1+(i/\omega \tau) \chi^0(q,\omega+i/\tau) /\chi^0(q,0) }.
\label{eq:NonintResp}
\end{equation}
For the TI modeled in our study, we choose a relaxation time of $\tau=500$ fs.\cite{Dong_Science, Qi_APL, Glinka_APL, Sobota_PRL} Within the random phase approximation, the response function $\chi^0_{\tau}(q,\omega)$ has the same expression as in the case of graphene,\cite{Wunsch.NJP, Hwang_PRB_2007} but without degeneracy neither in valley nor in spin.

We describe next how to get the dispersion relationship directly from this many-body approach, for a general system with an arbitrary number of   \fermionlayers\ that interact via Coulomb coupling. In contrast to the classical electrodynamic formalism used in much of this work, where the interaction between \fermionlayers\ is captured via the full Maxwell's equations, the many-body  approach ignores retardation of the fields, which should be valid for the extremely thin systems under consideration.

 Each \fermionlayer\ is surrounded by dielectric media with certain bulk dielectric constant and it is localized in the vertical $z$-direction, with the localization $|\lambda_i(z)|$ (where $i$ runs over the layers) given by
\begin{equation}\label{sin_sqr_local}
|\lambda_i(z)|^2=\frac{2}{h_i}\sin^2\left(\frac{\pi (z^0_i-z)}{h_i}\right)\theta(h_i-(z^0_i-z))\theta(z^0_i-z).
\end{equation}
$h_i$ is the width of the respective \fermionlayer\ in the $z$ direction, while $z^0_i$ sets its upper plane. The width $h_i$ accounts for the extension of the TI surface state in the direction normal to the surface plane.

The non-interacting response function of the \fermionlayers\ as a function of wavenumber reads\cite{eremeev15}
\begin{equation}\label{chi_0_sum}
\chi^0_{\tau}(q,\omega;z,z')=\sum\limits_{i} |\lambda_i(z)|^2 \chi^0_{\tau}(q,\omega) |\lambda_i(z')|^2,
\end{equation}
As a result, the interacting response function can be found as
\begin{equation}\label{chi_expanded}
\chi_{\tau}(q,\omega;z,z')=\sum\limits_{i,j} |\lambda_i(z)|^2 \chi_{ij}(q,\omega) |\lambda_j(z')|^2,
\end{equation}
 where
\begin{equation}\label{chi_2D_series}
\chi_{ij}(q,\omega)=\chi^0_{\tau}(q,\omega)\delta_{ij}+\chi^0_{\tau}( q,\omega)W_{ij}(q,\omega)\chi^0_{\tau}(q,\omega)
\end{equation}
with the Kronecker delta $\delta_{ij}$ and the screened interaction $W_{ij}(q)$ between $i-$th and $j-$th \fermionlayers\ defined as
\begin{equation}\label{screened_interaction}
W_{ij}(q,\omega)=\left[\left(\mathbf{1}-\mathbf{U}\bm{\chi^0}_{\tau}\right)^{-1}\mathbf{U}\right]_{ij}.
\end{equation}
Here the matrix $\bm{\chi^0}_{\tau}$ is considered as a diagonal matrix with the elements $\chi^0_{\tau}\delta_{ij}$. The elements of the matrix \textbf{U} are given by
\begin{equation}\label{screened_int_elements}
\left[\mathbf{U}\right]_{ij}\equiv U_{ij}=\int dz dz' |\lambda_i(z)|^2 \phi(q;z, z') |\lambda_j(z')|^2.
\end{equation}
The function $\phi(q;z, z')$ entering this equation is the Coulomb interaction as obtained from the Poisson equation with the $z$-dependent dielectric constant $\varepsilon(z)$ corresponding to the considered geometry.  Finally, the energies of the collective excitations in the set of 2D-FLs are defined by locating the poles of $\chi_{\tau}$ of Eq. (\ref{chi_expanded}). We find these poles from the zero crossing of $\mathrm{det}|\mathbf{1}-\mathbf{U}\bm{\chi}_{\tau}^0|$. In the case of the semi-infinite substrate, we place only one \fermionlayer\ with a non-zero thickness $\sim 10$ $\ANGST$ \, such as its upper plane coincides with the TI surface. When considering the TI film, we add one more \fermionlayer\ of the same thickness and charge density on the other side of the film, where the lower plane of this layer is superimposed with the bottom of the TI film. Thus, the $10$\,nm thick TI insulator layer studied in the text includes the thickness of both \fermionlayers.

\section {Appendix B: Classical Calculations}
\label{COMSOLnonlocal}
\begin{figure}[tb!] 
\includegraphics[width=0.5 \textwidth]{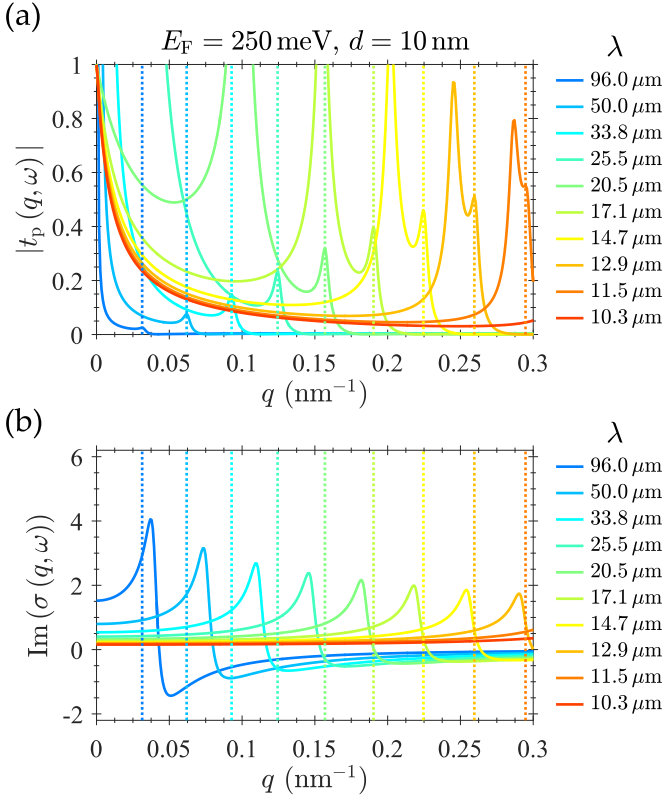}  \centering
\caption{(a) Dependence with parallel wavenumber of the absolute values of the transmission coefficient $|t_p(q,\omega)|$ characterizing a $10$\,nm thin TI layer, for several incident wavelengths. (b)  Imaginary part of the non-local conductivity $\sigma(q,\omega)$  as a function of the parallel wavenumber $q$  , plotted for the same incident wavelengths  as in (a). The dashed vertical lines in (a) and (b) are visual guides corresponding to the peaks in (a) associated with the acoustic plasmon. A second peak appearing in  the spectra in (a) at  lower  $q$ corresponds to the optical mode. The bulk dielectric constant of the TI is $\varepsilon_{\rm{b}}$=25 and  the Fermi energy $E_{\rm{F}}$=$250$\,meV  }
\label{nonlocal}
\end{figure}

We describe next in more details how the classical calculations are performed once $\sigma(q,\omega)$ is obtained from the many-body results. In the case of thin structures with two \fermionlayers, we assume\cite{Scharf_PRB_2012,Principi_PRB_2012} that the layers are sufficiently far from each other to neglect inter-layer transitions, so that we can directly use the conductivity $\sigma(q,\omega)$ obtained for a single interface as described in Appendix A. For our symmetric thin structures $\sigma(q,\omega)$ is the same for both TI-vacuum interfaces. The \fermionlayers\  interact via electromagnetic (Coulomb) coupling.

For these calculations, the bulk TI and the surrounding vacuum are characterized by their permittivity ($\varepsilon_1=1$ and $\varepsilon_{\rm{b}}=25$, respectively), so that we ignore any phononic mode. We also neglect the  correction to the electromagnetic constitutive  equations  due to induced currents at the \fermionlayers\ \cite{qi11,karch11}  as we expect that this correction will only introduce second-order effects to the main phenomena studied here\cite{QiPRB,SchutkyPRB}.


For the calculation of the response of the  substrate and thin film (Fig. \ref{schematic_1}(a,b)), which are illuminated by a  point dipole, we use the plane-wave decomposition method, where the electromagnetic fields of the dipolar source are decomposed as  infinitely many plane waves of different wavenumbers\cite{novotnyBook,Aiz_sub_enh}.

The induced fields  $E^{\rm{ind}}$ result from an integral over the response to each of these plane waves, as calculated from the reflection and transmission coefficients at each interface, together with the propagation in the TI and surrounding vacuum. For the calculation, it is important to take into account that the  \fermionlayers\ modify the typical Fresnel reflection and transmission coefficients  \cite{koppens_nl}. To illustrate the influence of the \fermionlayers\ on these coefficients, we give next the reflection $r_{{\rm{TI}}}$ between vacuum and the TI for a plane-wave with vacuum wavenumber $k_1$ ($k_{\rm{TI}}$ in the TI) and parallel wavenumber $q$

\begin{equation}
r_{{\rm{TI}}}(q,\omega)  = \frac{  \varepsilon_{\rm{TI}}  k^1_{z} - \varepsilon_1 k^{\rm{TI}}_{z} + \frac{4\pi\sigma k^1_{z} k^{\rm{TI}}_{z}}{ \omega}} {  \varepsilon_{\rm{TI}}  k^1_{z} + \varepsilon_1 k^{\rm{TI}}_{z} + \frac{4\pi\sigma k^1_{z} k^{\rm{TI}}_{z}}{ \omega}}
\label{ref_substrate}
\end{equation}

where $k^1_{z}=\sqrt{k^2_1-q^2}$ and $k^{\rm{TI}}_{z}=\sqrt{k^2_{\rm{TI}}-q^2}$ are the normal components  of the plane-wave at both sides of the interface.  This calculation directly takes non-locality into account by adopting the corresponding conductivity $\sigma(q,\omega)$ for each plane-wave of parallel wavenumber $q$.

Besides giving the fields at all positions, the plane-wave decomposition allows to obtain the dispersion-relationship from the position of the resonances, which we then can compare with the result from the many-body calculations.  More exactly,  we calculate the dispersion relationship from the classical approach by calculating for each frequency of interest the parallel plasmon wavenumber $q_{\rm{\rm{SP}}}$ that results in a larger transmission $t_p(q,\omega)$  of the full system. With this approach it is also possible  to calculate the dispersion for the often-used local approximation, by simply using the conductivity  $\sigma(q=0,\omega)$ for all plane-waves.

In the case of the full-wave simulations of the disks\cite{comsol}, it is possible to include a 2D-layer at both the upper and bottom interfaces. However, the conductivity $\sigma$ only depends on $\omega$. We are able to introduce  the effect of non-locality in an approximate manner because, for the situations under study, the optical response  should be mostly determined by the response at one particular plasmonic parallel wavenumber $q_{\rm{SP}}$. $q_{\rm{SP}}$ can correspond to the parallel wavenumber of either the optical  ($q_{\rm{SP}}^{\rm{op}}$) or acoustic ($q_{\rm{SP}}^{\rm{ac}}$ ) modes (obtained from the dispersion relationship of the thin film), so that the calculations use $\sigma(q_{\rm{SP}}^{\rm{op}},\omega)$ or $\sigma(q_{\rm{SP}}^{\rm{ac}},\omega)$. For simplicity, we only consider the 2D-layer at the flat top and bottom interfaces, and not at the lateral side surfaces. The latter correspond to the so-called side faces (other than the (111) surface) of Bi$_2$Se$_3$ (a strong TI), which have surface state bands characterized by tilted anisotropic Dirac cones with notably reduced Fermi velocities and by spin-orbital texture than entirely differs from that of the (111) surface\cite{PhysRevB.93.085122, PhysRevB.84.195425}. As a consequence, the disk lateral sides would be described by a different conductivity and a possibly shorter relaxation time. To take into account the specific characteristics of the surface states on strong TI side faces, a proper approximation to the non-interacting response function should be developed first. However, we expect that it may introduce a significant shift on the peak position but should not affect the main physics described in this paper.

We examine next in more detail our approach to introduce non-locality into full-wave simulations. According to the discussion in the text, the localized acoustic and optical modes in the disks can be understood in terms of the excitation of  propagating plasmons, which get reflected at the edges and lead to Fabry-P\'erot-like resonances. Thus, to better  understand which wavenumbers dominate the response of the disk, we examine the behavior in $q$-space of the thin films.

With this purpose, we plot in  Fig. \ref{nonlocal}(a) the $q-$dependence of the absolute value of the transmission coefficient for the 10 nm thin TI disks, for different illumination wavelengths. Each spectra displays two peaks, corresponding to the acoustic mode at the largest $q$, and the optical at the lowest $q$.  If the following two conditions are verified, we should be able to use in the full-wave calculations the conductivity at either $q_{\rm{SP}}^{\rm{op}}$ or $q_{\rm{SP}}^{\rm{ac}}$ : i) the response should be dominated by the $q$-components of one of the peaks,  and ii) $\sigma(q)$ should not change strongly over the width of the peak.

To fulfill the first condition, we either use a  plane-wave to illuminate the structure, which does not excite the acoustic modes (large disks simulations), or we choose a frequency region where optical modes are not resonantly excited (calculations of the smaller disks).

To study in more detail the second condition, Fig. \ref{nonlocal}(b) shows the behavior of the imaginary part of $\sigma(q,\omega)$ for the same excitation wavelengths as in Fig. \ref{nonlocal}(a) (the real part gives similar conclusions). In general, the spectra varies relatively slowly with the parallel wavenumber, except for the presence of a clear resonant feature at a $q$ that depends on the excitation frequency and that is related to the edge of the electron-hole continuum (smoothed in the presence of non-zero losses). For excitation wavelengths  $\lambda\lesssim 50~\mu$m, the relatively flat region of the conductivity extend over the $q$-values corresponding to both the acoustic and optical peaks in \ref{nonlocal}(a). Thus, we expect our results in Fig. \ref{fig:ff_nf_disk600}, \ref{fig:ff_nf_disk50} to be a reasonable approximation of the non-local response.


%


\begin{thebibliography}{94}
\expandafter\ifx\csname natexlab\endcsname\relax\def\natexlab#1{#1}\fi
\expandafter\ifx\csname bibnamefont\endcsname\relax
  \def\bibnamefont#1{#1}\fi
\expandafter\ifx\csname bibfnamefont\endcsname\relax
  \def\bibfnamefont#1{#1}\fi
\expandafter\ifx\csname citenamefont\endcsname\relax
  \def\citenamefont#1{#1}\fi
\expandafter\ifx\csname url\endcsname\relax
  \def\url#1{\texttt{#1}}\fi
\expandafter\ifx\csname urlprefix\endcsname\relax\def\urlprefix{URL }\fi
\providecommand{\bibinfo}[2]{#2}
\providecommand{\eprint}[2][]{\url{#2}}

\bibitem[{\citenamefont{Barnes et~al.}(2003)\citenamefont{Barnes, Dereux, and
  Ebbesen}}]{barnes2003}
\bibinfo{author}{\bibfnamefont{W.~L.} \bibnamefont{Barnes}},
  \bibinfo{author}{\bibfnamefont{A.}~\bibnamefont{Dereux}}, \bibnamefont{and}
  \bibinfo{author}{\bibfnamefont{T.~W.} \bibnamefont{Ebbesen}},
  \bibinfo{journal}{Nature} \textbf{\bibinfo{volume}{424}},
  \bibinfo{pages}{824} (\bibinfo{year}{2003}).

\bibitem[{\citenamefont{Pitarke et~al.}(2006)\citenamefont{Pitarke, Silkin,
  Chulkov, and Echenique}}]{pitarke2006}
\bibinfo{author}{\bibfnamefont{J.}~\bibnamefont{Pitarke}},
  \bibinfo{author}{\bibfnamefont{V.}~\bibnamefont{Silkin}},
  \bibinfo{author}{\bibfnamefont{E.}~\bibnamefont{Chulkov}}, \bibnamefont{and}
  \bibinfo{author}{\bibfnamefont{P.}~\bibnamefont{Echenique}},
  \bibinfo{journal}{Reports on progress in physics}
  \textbf{\bibinfo{volume}{70}}, \bibinfo{pages}{1} (\bibinfo{year}{2006}).

\bibitem[{\citenamefont{Pelton et~al.}(2008)\citenamefont{Pelton, Aizpurua, and
  Bryant}}]{pelton2008}
\bibinfo{author}{\bibfnamefont{M.}~\bibnamefont{Pelton}},
  \bibinfo{author}{\bibfnamefont{J.}~\bibnamefont{Aizpurua}}, \bibnamefont{and}
  \bibinfo{author}{\bibfnamefont{G.}~\bibnamefont{Bryant}},
  \bibinfo{journal}{Laser \& Photonics Reviews} \textbf{\bibinfo{volume}{2}},
  \bibinfo{pages}{136} (\bibinfo{year}{2008}).

\bibitem[{\citenamefont{Orlita and Potemski}(2010)}]{DPlasmons}
\bibinfo{author}{\bibfnamefont{M.}~\bibnamefont{Orlita}} \bibnamefont{and}
  \bibinfo{author}{\bibfnamefont{M.}~\bibnamefont{Potemski}},
  \bibinfo{journal}{Semicond. Sci. Technol.} \textbf{\bibinfo{volume}{25}},
  \bibinfo{pages}{063001} (\bibinfo{year}{2010}).

\bibitem[{\citenamefont{Ju et~al.}(2011)\citenamefont{Ju, Baisong, Horng,
  Girit, Martin, Hao, Bechtel, Liang, Zettl, Shen et~al.}}]{Gr.THz.NatNano}
\bibinfo{author}{\bibfnamefont{L.}~\bibnamefont{Ju}},
  \bibinfo{author}{\bibnamefont{Baisong}},
  \bibinfo{author}{\bibfnamefont{J.}~\bibnamefont{Horng}},
  \bibinfo{author}{\bibfnamefont{C.}~\bibnamefont{Girit}},
  \bibinfo{author}{\bibfnamefont{M.}~\bibnamefont{Martin}},
  \bibinfo{author}{\bibfnamefont{Z.}~\bibnamefont{Hao}},
  \bibinfo{author}{\bibfnamefont{H.~A.} \bibnamefont{Bechtel}},
  \bibinfo{author}{\bibfnamefont{X.}~\bibnamefont{Liang}},
  \bibinfo{author}{\bibfnamefont{A.}~\bibnamefont{Zettl}},
  \bibinfo{author}{\bibfnamefont{Y.~R.} \bibnamefont{Shen}},
  \bibnamefont{et~al.}, \bibinfo{journal}{Nat. Nanotechnol.}
  \textbf{\bibinfo{volume}{6}}, \bibinfo{pages}{630} (\bibinfo{year}{2011}).

\bibitem[{\citenamefont{Nikitin et~al.}(2016)\citenamefont{Nikitin,
  Alonso-Gonz{\'a}lez, V{\'e}lez, Mastel, Centeno, Pesquera, Zurutuza,
  Casanova, Hueso, Koppens et~al.}}]{nikitin16}
\bibinfo{author}{\bibfnamefont{A.}~\bibnamefont{Nikitin}},
  \bibinfo{author}{\bibfnamefont{P.}~\bibnamefont{Alonso-Gonz{\'a}lez}},
  \bibinfo{author}{\bibfnamefont{S.}~\bibnamefont{V{\'e}lez}},
  \bibinfo{author}{\bibfnamefont{S.}~\bibnamefont{Mastel}},
  \bibinfo{author}{\bibfnamefont{A.}~\bibnamefont{Centeno}},
  \bibinfo{author}{\bibfnamefont{A.}~\bibnamefont{Pesquera}},
  \bibinfo{author}{\bibfnamefont{A.}~\bibnamefont{Zurutuza}},
  \bibinfo{author}{\bibfnamefont{F.}~\bibnamefont{Casanova}},
  \bibinfo{author}{\bibfnamefont{L.}~\bibnamefont{Hueso}},
  \bibinfo{author}{\bibfnamefont{F.}~\bibnamefont{Koppens}},
  \bibnamefont{et~al.}, \bibinfo{journal}{Nat. Photon.}
  \textbf{\bibinfo{volume}{10}}, \bibinfo{pages}{239} (\bibinfo{year}{2016}).

\bibitem[{\citenamefont{Fang et~al.}(2013)\citenamefont{Fang, Wang, Schlather,
  Liu, Ajayan, García~de Abajo, Nordlander, Zhu, and Halas}}]{fang13}
\bibinfo{author}{\bibfnamefont{Z.}~\bibnamefont{Fang}},
  \bibinfo{author}{\bibfnamefont{Y.}~\bibnamefont{Wang}},
  \bibinfo{author}{\bibfnamefont{A.~E.} \bibnamefont{Schlather}},
  \bibinfo{author}{\bibfnamefont{Z.}~\bibnamefont{Liu}},
  \bibinfo{author}{\bibfnamefont{P.~M.} \bibnamefont{Ajayan}},
  \bibinfo{author}{\bibfnamefont{F.~J.} \bibnamefont{García~de Abajo}},
  \bibinfo{author}{\bibfnamefont{P.}~\bibnamefont{Nordlander}},
  \bibinfo{author}{\bibfnamefont{X.}~\bibnamefont{Zhu}}, \bibnamefont{and}
  \bibinfo{author}{\bibfnamefont{N.~J.} \bibnamefont{Halas}},
  \bibinfo{journal}{Nano lett.} \textbf{\bibinfo{volume}{14}},
  \bibinfo{pages}{299} (\bibinfo{year}{2013}).

\bibitem[{\citenamefont{Yan et~al.}(2012)\citenamefont{Yan, Li, Chandra,
  Tulevski, Wu, Freitag, Zhu, Avouris, and Xia}}]{yan12}
\bibinfo{author}{\bibfnamefont{H.}~\bibnamefont{Yan}},
  \bibinfo{author}{\bibfnamefont{X.}~\bibnamefont{Li}},
  \bibinfo{author}{\bibfnamefont{B.}~\bibnamefont{Chandra}},
  \bibinfo{author}{\bibfnamefont{G.}~\bibnamefont{Tulevski}},
  \bibinfo{author}{\bibfnamefont{Y.}~\bibnamefont{Wu}},
  \bibinfo{author}{\bibfnamefont{M.}~\bibnamefont{Freitag}},
  \bibinfo{author}{\bibfnamefont{W.}~\bibnamefont{Zhu}},
  \bibinfo{author}{\bibfnamefont{P.}~\bibnamefont{Avouris}}, \bibnamefont{and}
  \bibinfo{author}{\bibfnamefont{F.}~\bibnamefont{Xia}},
  \bibinfo{journal}{Nature nanotechnol.} \textbf{\bibinfo{volume}{7}},
  \bibinfo{pages}{330} (\bibinfo{year}{2012}).

\bibitem[{\citenamefont{Mikhailov}(2011)}]{mikhailov11}
\bibinfo{author}{\bibfnamefont{S.~A.} \bibnamefont{Mikhailov}},
  \bibinfo{journal}{Phys. Rev. B} \textbf{\bibinfo{volume}{84}},
  \bibinfo{pages}{045432} (\bibinfo{year}{2011}).

\bibitem[{\citenamefont{Garc\'ia~de Abajo}(2014)}]{G.Abajo.Gr.Review}
\bibinfo{author}{\bibfnamefont{F.~J.} \bibnamefont{Garc\'ia~de Abajo}},
  \bibinfo{journal}{ACS Photonics} \textbf{\bibinfo{volume}{1}},
  \bibinfo{pages}{135} (\bibinfo{year}{2014}).

\bibitem[{\citenamefont{Koppens et~al.}(2011)\citenamefont{Koppens, Chang, and
  Garc\'{i}a~de Abajo}}]{koppens_nl}
\bibinfo{author}{\bibfnamefont{F.~H.~L.} \bibnamefont{Koppens}},
  \bibinfo{author}{\bibfnamefont{D.~E.} \bibnamefont{Chang}}, \bibnamefont{and}
  \bibinfo{author}{\bibfnamefont{F.~J.} \bibnamefont{Garc\'{i}a~de Abajo}},
  \bibinfo{journal}{Nano Lett.} \textbf{\bibinfo{volume}{11}},
  \bibinfo{pages}{3370} (\bibinfo{year}{2011}).

\bibitem[{\citenamefont{Chen et~al.}(2012)\citenamefont{Chen, Badioli,
  Alonso-Gonzalez, Thongrattanasiri, Osmond, Spasenovic, Centeno, Pesquera,
  Godignon, Elorza et~al.}}]{jianing_nature}
\bibinfo{author}{\bibfnamefont{J.}~\bibnamefont{Chen}},
  \bibinfo{author}{\bibfnamefont{M.}~\bibnamefont{Badioli}},
  \bibinfo{author}{\bibfnamefont{P.}~\bibnamefont{Alonso-Gonzalez}},
  \bibinfo{author}{\bibfnamefont{S.~F.~H.} \bibnamefont{Thongrattanasiri}},
  \bibinfo{author}{\bibfnamefont{J.}~\bibnamefont{Osmond}},
  \bibinfo{author}{\bibfnamefont{M.}~\bibnamefont{Spasenovic}},
  \bibinfo{author}{\bibfnamefont{A.}~\bibnamefont{Centeno}},
  \bibinfo{author}{\bibfnamefont{A.}~\bibnamefont{Pesquera}},
  \bibinfo{author}{\bibfnamefont{P.}~\bibnamefont{Godignon}},
  \bibinfo{author}{\bibfnamefont{A.~Z.} \bibnamefont{Elorza}},
  \bibnamefont{et~al.}, \bibinfo{journal}{Nature}
  \textbf{\bibinfo{volume}{487}}, \bibinfo{pages}{77 } (\bibinfo{year}{2012}).

\bibitem[{\citenamefont{Jablan et~al.}(2009)\citenamefont{Jablan, Buljan, and
  Soljacic}}]{JablanPRB}
\bibinfo{author}{\bibfnamefont{M.}~\bibnamefont{Jablan}},
  \bibinfo{author}{\bibfnamefont{H.}~\bibnamefont{Buljan}}, \bibnamefont{and}
  \bibinfo{author}{\bibfnamefont{M.}~\bibnamefont{Soljacic}},
  \bibinfo{journal}{Phys. Rev. B} \textbf{\bibinfo{volume}{80}},
  \bibinfo{pages}{245435} (\bibinfo{year}{2009}).

\bibitem[{\citenamefont{Fei et~al.}(2012)\citenamefont{Fei, Rodin, Andreev,
  Bao, McLeod, Wagner, Zhang, M., Zhao, Thiemens et~al.}}]{Basov_gr}
\bibinfo{author}{\bibfnamefont{Z.}~\bibnamefont{Fei}},
  \bibinfo{author}{\bibfnamefont{A.~S.} \bibnamefont{Rodin}},
  \bibinfo{author}{\bibfnamefont{G.~O.} \bibnamefont{Andreev}},
  \bibinfo{author}{\bibfnamefont{W.}~\bibnamefont{Bao}},
  \bibinfo{author}{\bibfnamefont{A.~S.} \bibnamefont{McLeod}},
  \bibinfo{author}{\bibfnamefont{M.}~\bibnamefont{Wagner}},
  \bibinfo{author}{\bibnamefont{Zhang}},
  \bibinfo{author}{\bibfnamefont{L.}~\bibnamefont{M.}},
  \bibinfo{author}{\bibfnamefont{Z.}~\bibnamefont{Zhao}},
  \bibinfo{author}{\bibfnamefont{M.}~\bibnamefont{Thiemens}},
  \bibnamefont{et~al.}, \bibinfo{journal}{Nature}
  \textbf{\bibinfo{volume}{487}}, \bibinfo{pages}{82} (\bibinfo{year}{2012}).

\bibitem[{\citenamefont{Manjavacas and Garc\'{i}a~de
  Abajo}(2013)}]{AuMono.G.Abajo}
\bibinfo{author}{\bibfnamefont{A.}~\bibnamefont{Manjavacas}} \bibnamefont{and}
  \bibinfo{author}{\bibfnamefont{F.~J.} \bibnamefont{Garc\'{i}a~de Abajo}},
  \bibinfo{journal}{Nat. Commun.} \textbf{\bibinfo{volume}{5}},
  \bibinfo{pages}{3548} (\bibinfo{year}{2014}).

\bibitem[{\citenamefont{Zhang and Zhang}(2009)}]{HaijunZhang}
\bibinfo{author}{\bibfnamefont{H.}~\bibnamefont{Zhang}} \bibnamefont{and}
  \bibinfo{author}{\bibfnamefont{S.~C.} \bibnamefont{Zhang}},
  \bibinfo{journal}{Nature Phys.} \textbf{\bibinfo{volume}{5}},
  \bibinfo{pages}{438 } (\bibinfo{year}{2009}).

\bibitem[{\citenamefont{Sch\"utky et~al.}(2013)\citenamefont{Sch\"utky, Ertler,
  Tr\"ugler, and Hohenester}}]{SchutkyPRB}
\bibinfo{author}{\bibfnamefont{R.}~\bibnamefont{Sch\"utky}},
  \bibinfo{author}{\bibfnamefont{C.}~\bibnamefont{Ertler}},
  \bibinfo{author}{\bibfnamefont{A.}~\bibnamefont{Tr\"ugler}},
  \bibnamefont{and}
  \bibinfo{author}{\bibfnamefont{U.}~\bibnamefont{Hohenester}},
  \bibinfo{journal}{Phys. Rev. B} \textbf{\bibinfo{volume}{88}},
  \bibinfo{pages}{195311} (\bibinfo{year}{2013}).

\bibitem[{\citenamefont{Hsieh and Hasan}(2009)}]{Hsieh-nature}
\bibinfo{author}{\bibfnamefont{D.}~\bibnamefont{Hsieh}} \bibnamefont{and}
  \bibinfo{author}{\bibfnamefont{M.~Z.} \bibnamefont{Hasan}},
  \bibinfo{journal}{Nature} \textbf{\bibinfo{volume}{460}},
  \bibinfo{pages}{1101} (\bibinfo{year}{2009}).

\bibitem[{\citenamefont{Lai et~al.}(2014)\citenamefont{Lai, Lin, Wu, and
  Liu}}]{Lai.TI.Plas}
\bibinfo{author}{\bibfnamefont{Y.~P.} \bibnamefont{Lai}},
  \bibinfo{author}{\bibfnamefont{I.~T.} \bibnamefont{Lin}},
  \bibinfo{author}{\bibfnamefont{K.~H.} \bibnamefont{Wu}}, \bibnamefont{and}
  \bibinfo{author}{\bibfnamefont{J.~M.} \bibnamefont{Liu}},
  \bibinfo{journal}{Nanomater. Nanotechnol.} \textbf{\bibinfo{volume}{4}},
  \bibinfo{pages}{13} (\bibinfo{year}{2014}).

\bibitem[{\citenamefont{Sim et~al.}(2015)\citenamefont{Sim, Jang, Koirala,
  Brahlek, Moon, Sung, Park, Cha, Oh, Jo et~al.}}]{sim15}
\bibinfo{author}{\bibfnamefont{S.}~\bibnamefont{Sim}},
  \bibinfo{author}{\bibfnamefont{H.}~\bibnamefont{Jang}},
  \bibinfo{author}{\bibfnamefont{N.}~\bibnamefont{Koirala}},
  \bibinfo{author}{\bibfnamefont{M.}~\bibnamefont{Brahlek}},
  \bibinfo{author}{\bibfnamefont{J.}~\bibnamefont{Moon}},
  \bibinfo{author}{\bibfnamefont{J.~H.} \bibnamefont{Sung}},
  \bibinfo{author}{\bibfnamefont{J.}~\bibnamefont{Park}},
  \bibinfo{author}{\bibfnamefont{S.}~\bibnamefont{Cha}},
  \bibinfo{author}{\bibfnamefont{S.}~\bibnamefont{Oh}},
  \bibinfo{author}{\bibfnamefont{M.-H.} \bibnamefont{Jo}},
  \bibnamefont{et~al.}, \bibinfo{journal}{Nat. Commun.}
  \textbf{\bibinfo{volume}{6}},  \bibinfo{pages}{8814} (\bibinfo{year}{2015}).

\bibitem[{\citenamefont{Hasan and Kane}(2010)}]{hasan_Colloquium}
\bibinfo{author}{\bibfnamefont{M.~Z.} \bibnamefont{Hasan}} \bibnamefont{and}
  \bibinfo{author}{\bibfnamefont{C.~L.} \bibnamefont{Kane}},
  \bibinfo{journal}{Rev. Mod. Phys.} \textbf{\bibinfo{volume}{82}},
  \bibinfo{pages}{3045} (\bibinfo{year}{2010}).

\bibitem[{\citenamefont{Kane and Mele}(2005)}]{Kane_Mele_PRL}
\bibinfo{author}{\bibfnamefont{C.~L.} \bibnamefont{Kane}} \bibnamefont{and}
  \bibinfo{author}{\bibfnamefont{E.~J.} \bibnamefont{Mele}},
  \bibinfo{journal}{Phys. Rev. Lett.} \textbf{\bibinfo{volume}{95}},
  \bibinfo{pages}{226801} (\bibinfo{year}{2005}).

\bibitem[{\citenamefont{Wu et~al.}(2006)\citenamefont{Wu, Bernevig, and
  Zhang}}]{Wu_PRL}
\bibinfo{author}{\bibfnamefont{C.}~\bibnamefont{Wu}},
  \bibinfo{author}{\bibfnamefont{B.~A.} \bibnamefont{Bernevig}},
  \bibnamefont{and} \bibinfo{author}{\bibfnamefont{S.~C.} \bibnamefont{Zhang}},
  \bibinfo{journal}{Phys. Rev. Lett.} \textbf{\bibinfo{volume}{96}},
  \bibinfo{pages}{106401} (\bibinfo{year}{2006}).

\bibitem[{\citenamefont{Lu et~al.}(2010)\citenamefont{Lu, Shan, Yao, Niu, and
  Shen}}]{PRB_Lu}
\bibinfo{author}{\bibfnamefont{H.~Z.} \bibnamefont{Lu}},
  \bibinfo{author}{\bibfnamefont{W.~Y.} \bibnamefont{Shan}},
  \bibinfo{author}{\bibfnamefont{W.}~\bibnamefont{Yao}},
  \bibinfo{author}{\bibfnamefont{Q.}~\bibnamefont{Niu}}, \bibnamefont{and}
  \bibinfo{author}{\bibfnamefont{S.~Q.} \bibnamefont{Shen}},
  \bibinfo{journal}{Phys. Rev. B} \textbf{\bibinfo{volume}{81}},
  \bibinfo{pages}{115407} (\bibinfo{year}{2010}).

\bibitem[{\citenamefont{Di~Pietro et~al.}(2013)\citenamefont{Di~Pietro,
  Ortolani, Limaj, Di~Gaspare, Giliberti, Giorgianni, Brahlek, Bansal, Koirala,
  Oh et~al.}}]{pietro_natnano13}
\bibinfo{author}{\bibfnamefont{P.}~\bibnamefont{Di~Pietro}},
  \bibinfo{author}{\bibfnamefont{M.}~\bibnamefont{Ortolani}},
  \bibinfo{author}{\bibfnamefont{O.}~\bibnamefont{Limaj}},
  \bibinfo{author}{\bibfnamefont{A.}~\bibnamefont{Di~Gaspare}},
  \bibinfo{author}{\bibfnamefont{V.}~\bibnamefont{Giliberti}},
  \bibinfo{author}{\bibnamefont{Giorgianni}},
  \bibinfo{author}{\bibnamefont{Brahlek}},
  \bibinfo{author}{\bibfnamefont{N.}~\bibnamefont{Bansal}},
  \bibinfo{author}{\bibfnamefont{N.}~\bibnamefont{Koirala}},
  \bibinfo{author}{\bibfnamefont{S.}~\bibnamefont{Oh}}, \bibnamefont{et~al.},
  \bibinfo{journal}{Nat. Nanotechnol.} \textbf{\bibinfo{volume}{8}},
  \bibinfo{pages}{556 } (\bibinfo{year}{2013}).

\bibitem[{\citenamefont{Ou et~al.}(2014)\citenamefont{Ou, So, Adamo, Sulaev,
  Wang, and Zheludev}}]{uv-vis-topo}
\bibinfo{author}{\bibfnamefont{J.~Y.} \bibnamefont{Ou}},
  \bibinfo{author}{\bibfnamefont{J.~K.} \bibnamefont{So}},
  \bibinfo{author}{\bibfnamefont{G.}~\bibnamefont{Adamo}},
  \bibinfo{author}{\bibfnamefont{A.}~\bibnamefont{Sulaev}},
  \bibinfo{author}{\bibfnamefont{L.}~\bibnamefont{Wang}}, \bibnamefont{and}
  \bibinfo{author}{\bibfnamefont{N.~I.} \bibnamefont{Zheludev}},
  \bibinfo{journal}{Nat. Commun.} \textbf{\bibinfo{volume}{5}},
  \bibinfo{pages}{5139} (\bibinfo{year}{2014}).

\bibitem[{\citenamefont{Autore et~al.}(2015{\natexlab{a}})\citenamefont{Autore,
  Engelkamp, DApuzzo, Gaspare, Pietro, Vecchio, Brahlek, Koirala, Oh, and
  Lupi}}]{autore15}
\bibinfo{author}{\bibfnamefont{M.}~\bibnamefont{Autore}},
  \bibinfo{author}{\bibfnamefont{H.}~\bibnamefont{Engelkamp}},
  \bibinfo{author}{\bibfnamefont{F.}~\bibnamefont{DApuzzo}},
  \bibinfo{author}{\bibfnamefont{A.~D.} \bibnamefont{Gaspare}},
  \bibinfo{author}{\bibfnamefont{P.~D.} \bibnamefont{Pietro}},
  \bibinfo{author}{\bibfnamefont{I.~L.} \bibnamefont{Vecchio}},
  \bibinfo{author}{\bibfnamefont{M.}~\bibnamefont{Brahlek}},
  \bibinfo{author}{\bibfnamefont{N.}~\bibnamefont{Koirala}},
  \bibinfo{author}{\bibfnamefont{S.}~\bibnamefont{Oh}}, \bibnamefont{and}
  \bibinfo{author}{\bibfnamefont{S.}~\bibnamefont{Lupi}}, \bibinfo{journal}{ACS
  Photonics} \textbf{\bibinfo{volume}{2}}, \bibinfo{pages}{1231}
  (\bibinfo{year}{2015}{\natexlab{a}}).

\bibitem[{\citenamefont{Deshko et~al.}(2016)\citenamefont{Deshko,
  Krusin-Elbaum, Menon, Khanikaev, and Trevino}}]{Deshko16}
\bibinfo{author}{\bibfnamefont{Y.}~\bibnamefont{Deshko}},
  \bibinfo{author}{\bibfnamefont{L.}~\bibnamefont{Krusin-Elbaum}},
  \bibinfo{author}{\bibfnamefont{V.}~\bibnamefont{Menon}},
  \bibinfo{author}{\bibfnamefont{A.}~\bibnamefont{Khanikaev}},
  \bibnamefont{and} \bibinfo{author}{\bibfnamefont{J.}~\bibnamefont{Trevino}},
  \bibinfo{journal}{Opt. Express} \textbf{\bibinfo{volume}{24}},
  \bibinfo{pages}{7398} (\bibinfo{year}{2016}).

\bibitem[{\citenamefont{Guozhi et~al.}(2016)\citenamefont{Guozhi, Peng,
  Yanbang, and Kai}}]{guozhi16}
\bibinfo{author}{\bibfnamefont{J.}~\bibnamefont{Guozhi}},
  \bibinfo{author}{\bibfnamefont{W.}~\bibnamefont{Peng}},
  \bibinfo{author}{\bibfnamefont{Z.}~\bibnamefont{Yanbang}}, \bibnamefont{and}
  \bibinfo{author}{\bibfnamefont{C.}~\bibnamefont{Kai}}, \bibinfo{journal}{Sci.
  Rep.} \textbf{\bibinfo{volume}{6}}, \bibinfo{pages}{25884}
  (\bibinfo{year}{2016}).

\bibitem[{\citenamefont{Raghu et~al.}(2010)\citenamefont{Raghu, Chung, Qi, and
  Zhang}}]{RaghuPRL}
\bibinfo{author}{\bibfnamefont{S.}~\bibnamefont{Raghu}},
  \bibinfo{author}{\bibfnamefont{S.~B.} \bibnamefont{Chung}},
  \bibinfo{author}{\bibfnamefont{X.~L.} \bibnamefont{Qi}}, \bibnamefont{and}
  \bibinfo{author}{\bibfnamefont{S.~C.} \bibnamefont{Zhang}},
  \bibinfo{journal}{Phys. Rev. Lett.} \textbf{\bibinfo{volume}{104}},
  \bibinfo{pages}{116401} (\bibinfo{year}{2010}).

\bibitem[{\citenamefont{Stauber}(2014)}]{Stauber14}
\bibinfo{author}{\bibfnamefont{T.}~\bibnamefont{Stauber}}, \bibinfo{journal}{J.
  Phys. Condens. Matter} \textbf{\bibinfo{volume}{26}}, \bibinfo{pages}{123201}
  (\bibinfo{year}{2014}).

\bibitem[{\citenamefont{Profumo et~al.}(2012)\citenamefont{Profumo, Asgari,
  Polini, and MacDonald}}]{profumo12}
\bibinfo{author}{\bibfnamefont{R.~E.~V.} \bibnamefont{Profumo}},
  \bibinfo{author}{\bibfnamefont{R.}~\bibnamefont{Asgari}},
  \bibinfo{author}{\bibfnamefont{M.}~\bibnamefont{Polini}}, \bibnamefont{and}
  \bibinfo{author}{\bibfnamefont{A.~H.} \bibnamefont{MacDonald}},
  \bibinfo{journal}{Phys. Rev. B} \textbf{\bibinfo{volume}{85}},
  \bibinfo{pages}{085443} (\bibinfo{year}{2012}).

\bibitem[{\citenamefont{Stauber and G{\'o}mez-Santos}(2012)}]{Stauber12c}
\bibinfo{author}{\bibfnamefont{T.}~\bibnamefont{Stauber}} \bibnamefont{and}
  \bibinfo{author}{\bibfnamefont{G.}~\bibnamefont{G{\'o}mez-Santos}},
  \bibinfo{journal}{New Journal of Physics} \textbf{\bibinfo{volume}{14}},
  \bibinfo{pages}{105018} (\bibinfo{year}{2012}).

\bibitem[{\citenamefont{Hwang and Das~Sarma}(2009)}]{hwang09}
\bibinfo{author}{\bibfnamefont{E.~H.} \bibnamefont{Hwang}} \bibnamefont{and}
  \bibinfo{author}{\bibfnamefont{S.}~\bibnamefont{Das~Sarma}},
  \bibinfo{journal}{Phys. Rev. B} \textbf{\bibinfo{volume}{80}},
  \bibinfo{pages}{205405} (\bibinfo{year}{2009}).

\bibitem[{\citenamefont{Stauber et~al.}(2013)\citenamefont{Stauber,
  G\'{o}mez-Santos, and Brey}}]{StauberPRB}
\bibinfo{author}{\bibfnamefont{T.}~\bibnamefont{Stauber}},
  \bibinfo{author}{\bibfnamefont{G.}~\bibnamefont{G\'{o}mez-Santos}},
  \bibnamefont{and} \bibinfo{author}{\bibfnamefont{L.}~\bibnamefont{Brey}},
  \bibinfo{journal}{Phys. Rev. B} \textbf{\bibinfo{volume}{88}},
  \bibinfo{pages}{205427} (\bibinfo{year}{2013}).

\bibitem[{\citenamefont{Stauber et~al.}(2017)\citenamefont{Stauber,
  G{\'o}mez-Santos, and Brey}}]{stauber17b}
\bibinfo{author}{\bibfnamefont{T.}~\bibnamefont{Stauber}},
  \bibinfo{author}{\bibfnamefont{G.}~\bibnamefont{G{\'o}mez-Santos}},
  \bibnamefont{and} \bibinfo{author}{\bibfnamefont{L.}~\bibnamefont{Brey}},
  \bibinfo{journal}{ACS Photonics} \textbf{\bibinfo{volume}{4}},
  \bibinfo{pages}{2978} (\bibinfo{year}{2017}).

\bibitem[{\citenamefont{Qu et~al.}(2010)\citenamefont{Qu, Hor, Xiong, Cava, and
  Ong}}]{Dong_Science}
\bibinfo{author}{\bibfnamefont{D.-X.} \bibnamefont{Qu}},
  \bibinfo{author}{\bibfnamefont{Y.~S.} \bibnamefont{Hor}},
  \bibinfo{author}{\bibfnamefont{J.}~\bibnamefont{Xiong}},
  \bibinfo{author}{\bibfnamefont{R.~J.} \bibnamefont{Cava}}, \bibnamefont{and}
  \bibinfo{author}{\bibfnamefont{N.~P.} \bibnamefont{Ong}},
  \bibinfo{journal}{Science} \textbf{\bibinfo{volume}{329}},
  \bibinfo{pages}{821} (\bibinfo{year}{2010}).

\bibitem[{\citenamefont{Qi et~al.}(2010)\citenamefont{Qi, Chen, Yu,
  Cadden-Zimansky, Smirnov, Tolk, Miotkowski, Cao, Chen, Wu et~al.}}]{Qi_APL}
\bibinfo{author}{\bibfnamefont{J.}~\bibnamefont{Qi}},
  \bibinfo{author}{\bibfnamefont{X.}~\bibnamefont{Chen}},
  \bibinfo{author}{\bibfnamefont{W.}~\bibnamefont{Yu}},
  \bibinfo{author}{\bibfnamefont{P.}~\bibnamefont{Cadden-Zimansky}},
  \bibinfo{author}{\bibfnamefont{D.}~\bibnamefont{Smirnov}},
  \bibinfo{author}{\bibfnamefont{N.~H.} \bibnamefont{Tolk}},
  \bibinfo{author}{\bibfnamefont{I.}~\bibnamefont{Miotkowski}},
  \bibinfo{author}{\bibfnamefont{H.}~\bibnamefont{Cao}},
  \bibinfo{author}{\bibfnamefont{Y.~P.} \bibnamefont{Chen}},
  \bibinfo{author}{\bibfnamefont{Y.}~\bibnamefont{Wu}}, \bibnamefont{et~al.},
  \bibinfo{journal}{Appl. Phys. Lett.} \textbf{\bibinfo{volume}{97}},
  \bibinfo{pages}{182102} (\bibinfo{year}{2010}).

\bibitem[{\citenamefont{Glinka et~al.}(2013)\citenamefont{Glinka, Babakiray,
  Johnson, Bristow, Holcomb, and Lederman}}]{Glinka_APL}
\bibinfo{author}{\bibfnamefont{Y.~D.} \bibnamefont{Glinka}},
  \bibinfo{author}{\bibfnamefont{S.}~\bibnamefont{Babakiray}},
  \bibinfo{author}{\bibfnamefont{T.~A.} \bibnamefont{Johnson}},
  \bibinfo{author}{\bibfnamefont{A.~D.} \bibnamefont{Bristow}},
  \bibinfo{author}{\bibfnamefont{M.~B.} \bibnamefont{Holcomb}},
  \bibnamefont{and} \bibinfo{author}{\bibfnamefont{D.}~\bibnamefont{Lederman}},
  \bibinfo{journal}{Appl. Phys. Lett.} \textbf{\bibinfo{volume}{103}},
  \bibinfo{pages}{151903} (\bibinfo{year}{2013}).

\bibitem[{\citenamefont{Sobota et~al.}(2012)\citenamefont{Sobota, Yang,
  Analytis, Chen, Fisher, Kirchmann, and Shen}}]{Sobota_PRL}
\bibinfo{author}{\bibfnamefont{J.~A.} \bibnamefont{Sobota}},
  \bibinfo{author}{\bibfnamefont{S.}~\bibnamefont{Yang}},
  \bibinfo{author}{\bibfnamefont{J.~G.} \bibnamefont{Analytis}},
  \bibinfo{author}{\bibfnamefont{Y.~L.} \bibnamefont{Chen}},
  \bibinfo{author}{\bibfnamefont{I.~R.} \bibnamefont{Fisher}},
  \bibinfo{author}{\bibfnamefont{P.~S.} \bibnamefont{Kirchmann}},
  \bibnamefont{and} \bibinfo{author}{\bibfnamefont{Z.~X.} \bibnamefont{Shen}},
  \bibinfo{journal}{Phys. Rev. Lett.} \textbf{\bibinfo{volume}{108}},
  \bibinfo{pages}{117403} (\bibinfo{year}{2012}).

\bibitem[{\citenamefont{Xia et~al.}(2009)\citenamefont{Xia, Qian, Hsieh, Wray,
  Pal, Lin, Bansil, Grauer, Hor, Cava et~al.}}]{xia2009observation}
\bibinfo{author}{\bibfnamefont{Y.}~\bibnamefont{Xia}},
  \bibinfo{author}{\bibfnamefont{D.}~\bibnamefont{Qian}},
  \bibinfo{author}{\bibfnamefont{D.}~\bibnamefont{Hsieh}},
  \bibinfo{author}{\bibfnamefont{L.}~\bibnamefont{Wray}},
  \bibinfo{author}{\bibfnamefont{A.}~\bibnamefont{Pal}},
  \bibinfo{author}{\bibfnamefont{H.}~\bibnamefont{Lin}},
  \bibinfo{author}{\bibfnamefont{A.}~\bibnamefont{Bansil}},
  \bibinfo{author}{\bibfnamefont{D.}~\bibnamefont{Grauer}},
  \bibinfo{author}{\bibfnamefont{Y.}~\bibnamefont{Hor}},
  \bibinfo{author}{\bibfnamefont{R.}~\bibnamefont{Cava}}, \bibnamefont{et~al.},
  \bibinfo{journal}{Nature Phys.} \textbf{\bibinfo{volume}{5}},
  \bibinfo{pages}{398} (\bibinfo{year}{2009}).

\bibitem[{\citenamefont{Nechaev et~al.}(2013)\citenamefont{Nechaev, Hatch,
  Bianchi, Guan, Friedrich, Aguilera, Mi, Iversen, Bl\"ugel, Hofmann
  et~al.}}]{IlyaPRB}
\bibinfo{author}{\bibfnamefont{I.~A.} \bibnamefont{Nechaev}},
  \bibinfo{author}{\bibfnamefont{R.~C.} \bibnamefont{Hatch}},
  \bibinfo{author}{\bibfnamefont{M.}~\bibnamefont{Bianchi}},
  \bibinfo{author}{\bibfnamefont{D.}~\bibnamefont{Guan}},
  \bibinfo{author}{\bibfnamefont{C.}~\bibnamefont{Friedrich}},
  \bibinfo{author}{\bibfnamefont{I.}~\bibnamefont{Aguilera}},
  \bibinfo{author}{\bibfnamefont{J.~L.} \bibnamefont{Mi}},
  \bibinfo{author}{\bibfnamefont{B.~B.} \bibnamefont{Iversen}},
  \bibinfo{author}{\bibfnamefont{S.}~\bibnamefont{Bl\"ugel}},
  \bibinfo{author}{\bibfnamefont{P.}~\bibnamefont{Hofmann}},
  \bibnamefont{et~al.}, \bibinfo{journal}{Phys. Rev. B}
  \textbf{\bibinfo{volume}{87}}, \bibinfo{pages}{121111}
  (\bibinfo{year}{2013}).

\bibitem[{\citenamefont{Stordeur et~al.}(1992)\citenamefont{Stordeur, Ketavong,
  Priemuth, Sobotta, and Riede}}]{PSSB.Stordeur}
\bibinfo{author}{\bibfnamefont{M.}~\bibnamefont{Stordeur}},
  \bibinfo{author}{\bibfnamefont{K.~K.} \bibnamefont{Ketavong}},
  \bibinfo{author}{\bibfnamefont{A.}~\bibnamefont{Priemuth}},
  \bibinfo{author}{\bibfnamefont{H.}~\bibnamefont{Sobotta}}, \bibnamefont{and}
  \bibinfo{author}{\bibfnamefont{V.}~\bibnamefont{Riede}},
  \bibinfo{journal}{Phys. Status Solidi (b)} \textbf{\bibinfo{volume}{169}},
  \bibinfo{pages}{505} (\bibinfo{year}{1992}).

\bibitem[{\citenamefont{Kogar et~al.}(2015)\citenamefont{Kogar, Vig, Thaler,
  Wong, Xiao, Reig-i Plessis, Cho, Valla, Pan, Schneeloch
  et~al.}}]{Kogar_2015arXiv}
\bibinfo{author}{\bibfnamefont{A.}~\bibnamefont{Kogar}},
  \bibinfo{author}{\bibfnamefont{S.}~\bibnamefont{Vig}},
  \bibinfo{author}{\bibfnamefont{A.}~\bibnamefont{Thaler}},
  \bibinfo{author}{\bibfnamefont{M.~H.} \bibnamefont{Wong}},
  \bibinfo{author}{\bibfnamefont{Y.}~\bibnamefont{Xiao}},
  \bibinfo{author}{\bibfnamefont{D.}~\bibnamefont{Reig-i Plessis}},
  \bibinfo{author}{\bibfnamefont{G.~Y.} \bibnamefont{Cho}},
  \bibinfo{author}{\bibfnamefont{T.}~\bibnamefont{Valla}},
  \bibinfo{author}{\bibfnamefont{Z.}~\bibnamefont{Pan}},
  \bibinfo{author}{\bibfnamefont{J.}~\bibnamefont{Schneeloch}},
  \bibnamefont{et~al.}, \bibinfo{journal}{Phys. Rev. Lett.}
  \textbf{\bibinfo{volume}{115}}, \bibinfo{pages}{257402}
  (\bibinfo{year}{2015}).

\bibitem[{\citenamefont{Eddrief et~al.}(2016)\citenamefont{Eddrief, Vidal, and
  Gallas}}]{Bi2Se3_diel}
\bibinfo{author}{\bibfnamefont{M.}~\bibnamefont{Eddrief}},
  \bibinfo{author}{\bibfnamefont{F.}~\bibnamefont{Vidal}}, \bibnamefont{and}
  \bibinfo{author}{\bibfnamefont{B.}~\bibnamefont{Gallas}},
  \bibinfo{journal}{Journal of Physics D: Applied Physics}
  \textbf{\bibinfo{volume}{49}}, \bibinfo{pages}{505304}
  (\bibinfo{year}{2016}).

\bibitem[{\citenamefont{Autore et~al.}(2015{\natexlab{b}})\citenamefont{Autore,
  D'Apuzzo, Di~Gaspare, Giliberti, Limaj, Roy, Brahlek, Koirala, Oh,
  Garc{\'\i}a~de Abajo et~al.}}]{autore15b}
\bibinfo{author}{\bibfnamefont{M.}~\bibnamefont{Autore}},
  \bibinfo{author}{\bibfnamefont{F.}~\bibnamefont{D'Apuzzo}},
  \bibinfo{author}{\bibfnamefont{A.}~\bibnamefont{Di~Gaspare}},
  \bibinfo{author}{\bibfnamefont{V.}~\bibnamefont{Giliberti}},
  \bibinfo{author}{\bibfnamefont{O.}~\bibnamefont{Limaj}},
  \bibinfo{author}{\bibfnamefont{P.}~\bibnamefont{Roy}},
  \bibinfo{author}{\bibfnamefont{M.}~\bibnamefont{Brahlek}},
  \bibinfo{author}{\bibfnamefont{N.}~\bibnamefont{Koirala}},
  \bibinfo{author}{\bibfnamefont{S.}~\bibnamefont{Oh}},
  \bibinfo{author}{\bibfnamefont{F.~J.} \bibnamefont{Garc{\'\i}a~de Abajo}},
  \bibnamefont{et~al.}, \bibinfo{journal}{Adv. Opt. Mater}
  \textbf{\bibinfo{volume}{3}}, \bibinfo{pages}{1257}
  (\bibinfo{year}{2015}{\natexlab{b}}).

\bibitem[{\citenamefont{Qi and Zhang}(2011)}]{qi11}
\bibinfo{author}{\bibfnamefont{X.-L.} \bibnamefont{Qi}} \bibnamefont{and}
  \bibinfo{author}{\bibfnamefont{S.-C.} \bibnamefont{Zhang}},
  \bibinfo{journal}{Rev. Mod. Phys.} \textbf{\bibinfo{volume}{83}},
  \bibinfo{pages}{1057} (\bibinfo{year}{2011}).

\bibitem[{\citenamefont{Karch}(2011)}]{karch11}
\bibinfo{author}{\bibfnamefont{A.}~\bibnamefont{Karch}},
  \bibinfo{journal}{Phys. Rev. B} \textbf{\bibinfo{volume}{83}},
  \bibinfo{pages}{245432} (\bibinfo{year}{2011}).

\bibitem[{\citenamefont{Feibelman}(1975)}]{feibelman75}
\bibinfo{author}{\bibfnamefont{P.~J.} \bibnamefont{Feibelman}},
  \bibinfo{journal}{Phys. Rev. B} \textbf{\bibinfo{volume}{12}},
  \bibinfo{pages}{1319} (\bibinfo{year}{1975}).

\bibitem[{\citenamefont{Liebsch and Schaich}(1995)}]{liebsch95}
\bibinfo{author}{\bibfnamefont{A.}~\bibnamefont{Liebsch}} \bibnamefont{and}
  \bibinfo{author}{\bibfnamefont{W.~L.} \bibnamefont{Schaich}},
  \bibinfo{journal}{Phys. Rev. B} \textbf{\bibinfo{volume}{52}},
  \bibinfo{pages}{14219} (\bibinfo{year}{1995}).

\bibitem[{\citenamefont{Ruppin}(1976)}]{ruppin76}
\bibinfo{author}{\bibfnamefont{R.}~\bibnamefont{Ruppin}}, \bibinfo{journal}{J.
  Opt. Soc. Am.} \textbf{\bibinfo{volume}{66}}, \bibinfo{pages}{449}
  (\bibinfo{year}{1976}).

\bibitem[{\citenamefont{Fuchs and Claro}(1987)}]{fuchs87}
\bibinfo{author}{\bibfnamefont{R.}~\bibnamefont{Fuchs}} \bibnamefont{and}
  \bibinfo{author}{\bibfnamefont{F.}~\bibnamefont{Claro}},
  \bibinfo{journal}{Phys. Rev. B} \textbf{\bibinfo{volume}{35}},
  \bibinfo{pages}{3722} (\bibinfo{year}{1987}).

\bibitem[{com()}]{comsol}
\bibinfo{note}{These calculations were performed using the finite element
  method (in frequency domain), using the RF module of the commercial software
  COMSOL Multiphysics}.

\bibitem[{\citenamefont{Wang et~al.}(2012)\citenamefont{Wang, Xiu, Cheng, He,
  Lang, Tang, Kou, Yu, Jiang, Chen et~al.}}]{Wang_NLett_gating}
\bibinfo{author}{\bibfnamefont{Y.}~\bibnamefont{Wang}},
  \bibinfo{author}{\bibfnamefont{F.}~\bibnamefont{Xiu}},
  \bibinfo{author}{\bibfnamefont{L.}~\bibnamefont{Cheng}},
  \bibinfo{author}{\bibfnamefont{L.}~\bibnamefont{He}},
  \bibinfo{author}{\bibfnamefont{M.}~\bibnamefont{Lang}},
  \bibinfo{author}{\bibfnamefont{J.}~\bibnamefont{Tang}},
  \bibinfo{author}{\bibfnamefont{X.}~\bibnamefont{Kou}},
  \bibinfo{author}{\bibfnamefont{X.}~\bibnamefont{Yu}},
  \bibinfo{author}{\bibfnamefont{X.}~\bibnamefont{Jiang}},
  \bibinfo{author}{\bibfnamefont{Z.}~\bibnamefont{Chen}}, \bibnamefont{et~al.},
  \bibinfo{journal}{Nano Lett.} \textbf{\bibinfo{volume}{12}},
  \bibinfo{pages}{1170} (\bibinfo{year}{2012}).

\bibitem[{\citenamefont{Xiu et~al.}(2011)\citenamefont{Xiu, He, Wang, Cheng,
  Chang, Lang, Huang, Kou, Zhou, Jiang et~al.}}]{Faxian-Nat.nano_gating}
\bibinfo{author}{\bibfnamefont{F.}~\bibnamefont{Xiu}},
  \bibinfo{author}{\bibfnamefont{L.}~\bibnamefont{He}},
  \bibinfo{author}{\bibfnamefont{Y.}~\bibnamefont{Wang}},
  \bibinfo{author}{\bibfnamefont{L.}~\bibnamefont{Cheng}},
  \bibinfo{author}{\bibfnamefont{L.-T.} \bibnamefont{Chang}},
  \bibinfo{author}{\bibfnamefont{M.}~\bibnamefont{Lang}},
  \bibinfo{author}{\bibfnamefont{G.}~\bibnamefont{Huang}},
  \bibinfo{author}{\bibfnamefont{X.}~\bibnamefont{Kou}},
  \bibinfo{author}{\bibfnamefont{Y.}~\bibnamefont{Zhou}},
  \bibinfo{author}{\bibfnamefont{X.}~\bibnamefont{Jiang}},
  \bibnamefont{et~al.}, \bibinfo{journal}{Nat. Nanotechnol.}
  \textbf{\bibinfo{volume}{6}}, \bibinfo{pages}{216} (\bibinfo{year}{2011}).

\bibitem[{\citenamefont{Otto}(1968)}]{otto68}
\bibinfo{author}{\bibfnamefont{A.}~\bibnamefont{Otto}},
  \bibinfo{journal}{Zeitschrift f{\"u}r Physik} \textbf{\bibinfo{volume}{216}},
  \bibinfo{pages}{398} (\bibinfo{year}{1968}).

\bibitem[{\citenamefont{Rindzevicius et~al.}(2007)\citenamefont{Rindzevicius,
  Alaverdyan, Sepulveda, Pakizeh, Kall, Hillenbrand, Aizpurua, and Garc\'ia~de
  Abajo}}]{Nanohole.Plasmons}
\bibinfo{author}{\bibfnamefont{T.}~\bibnamefont{Rindzevicius}},
  \bibinfo{author}{\bibfnamefont{Y.}~\bibnamefont{Alaverdyan}},
  \bibinfo{author}{\bibfnamefont{B.}~\bibnamefont{Sepulveda}},
  \bibinfo{author}{\bibfnamefont{T.}~\bibnamefont{Pakizeh}},
  \bibinfo{author}{\bibfnamefont{M.}~\bibnamefont{Kall}},
  \bibinfo{author}{\bibfnamefont{R.}~\bibnamefont{Hillenbrand}},
  \bibinfo{author}{\bibfnamefont{J.}~\bibnamefont{Aizpurua}}, \bibnamefont{and}
  \bibinfo{author}{\bibfnamefont{F.~J.} \bibnamefont{Garc\'ia~de Abajo}},
  \bibinfo{journal}{J. Phys. Chem. C} \textbf{\bibinfo{volume}{111}},
  \bibinfo{pages}{1207} (\bibinfo{year}{2007}).

\bibitem[{\citenamefont{Carminati et~al.}(1998)\citenamefont{Carminati,
  Nieto-Vesperinas, and Greffet}}]{carminati98}
\bibinfo{author}{\bibfnamefont{R.}~\bibnamefont{Carminati}},
  \bibinfo{author}{\bibfnamefont{M.}~\bibnamefont{Nieto-Vesperinas}},
  \bibnamefont{and} \bibinfo{author}{\bibfnamefont{J.-J.}
  \bibnamefont{Greffet}}, \bibinfo{journal}{J. Opt. Soc. Am. A}
  \textbf{\bibinfo{volume}{15}}, \bibinfo{pages}{706} (\bibinfo{year}{1998}).

\bibitem[{\citenamefont{Hanarp et~al.}(2003)\citenamefont{Hanarp, K{\"a}ll, and
  Sutherland}}]{hanarp03}
\bibinfo{author}{\bibfnamefont{P.}~\bibnamefont{Hanarp}},
  \bibinfo{author}{\bibfnamefont{M.}~\bibnamefont{K{\"a}ll}}, \bibnamefont{and}
  \bibinfo{author}{\bibfnamefont{D.~S.} \bibnamefont{Sutherland}},
  \bibinfo{journal}{J. Phys. Chem. B} \textbf{\bibinfo{volume}{107}},
  \bibinfo{pages}{5768} (\bibinfo{year}{2003}).

\bibitem[{\citenamefont{Esteban et~al.}(2008)\citenamefont{Esteban,
  Vogelgesang, Dorfmuller, Dmitriev, Rockstuhl, Etrich, and Kern}}]{esteban08}
\bibinfo{author}{\bibfnamefont{R.}~\bibnamefont{Esteban}},
  \bibinfo{author}{\bibfnamefont{R.}~\bibnamefont{Vogelgesang}},
  \bibinfo{author}{\bibfnamefont{J.}~\bibnamefont{Dorfmuller}},
  \bibinfo{author}{\bibfnamefont{A.}~\bibnamefont{Dmitriev}},
  \bibinfo{author}{\bibfnamefont{C.}~\bibnamefont{Rockstuhl}},
  \bibinfo{author}{\bibfnamefont{C.}~\bibnamefont{Etrich}}, \bibnamefont{and}
  \bibinfo{author}{\bibfnamefont{K.}~\bibnamefont{Kern}},
  \bibinfo{journal}{Nano Lett.} \textbf{\bibinfo{volume}{8}},
  \bibinfo{pages}{3155} (\bibinfo{year}{2008}).

\bibitem[{\citenamefont{Lassiter et~al.}(2010)\citenamefont{Lassiter, Sobhani,
  Fan, Kundu, Capasso, Nordlander, and Halas}}]{lassiter10}
\bibinfo{author}{\bibfnamefont{J.~B.} \bibnamefont{Lassiter}},
  \bibinfo{author}{\bibfnamefont{H.}~\bibnamefont{Sobhani}},
  \bibinfo{author}{\bibfnamefont{J.~A.} \bibnamefont{Fan}},
  \bibinfo{author}{\bibfnamefont{J.}~\bibnamefont{Kundu}},
  \bibinfo{author}{\bibfnamefont{F.}~\bibnamefont{Capasso}},
  \bibinfo{author}{\bibfnamefont{P.}~\bibnamefont{Nordlander}},
  \bibnamefont{and} \bibinfo{author}{\bibfnamefont{N.~J.} \bibnamefont{Halas}},
  \bibinfo{journal}{Nano Lett.} \textbf{\bibinfo{volume}{10}},
  \bibinfo{pages}{3184} (\bibinfo{year}{2010}).

\bibitem[{\citenamefont{Esteban et~al.}(2010)\citenamefont{Esteban, Teperik,
  and Greffet}}]{esteban10}
\bibinfo{author}{\bibfnamefont{R.}~\bibnamefont{Esteban}},
  \bibinfo{author}{\bibfnamefont{T.~V.} \bibnamefont{Teperik}},
  \bibnamefont{and} \bibinfo{author}{\bibfnamefont{J.~J.}
  \bibnamefont{Greffet}}, \bibinfo{journal}{Phys. Rev. Lett.}
  \textbf{\bibinfo{volume}{104}}, \bibinfo{pages}{026802}
  (\bibinfo{year}{2010}).

\bibitem[{\citenamefont{Bozhevolnyi and S{\o}ndergaard}(2007)}]{bozhevolnyi07}
\bibinfo{author}{\bibfnamefont{S.~I.} \bibnamefont{Bozhevolnyi}}
  \bibnamefont{and}
  \bibinfo{author}{\bibfnamefont{T.}~\bibnamefont{S{\o}ndergaard}},
  \bibinfo{journal}{Opt. Express} \textbf{\bibinfo{volume}{15}},
  \bibinfo{pages}{10869} (\bibinfo{year}{2007}).

\bibitem[{\citenamefont{Novotny}(2007)}]{novotny07}
\bibinfo{author}{\bibfnamefont{L.}~\bibnamefont{Novotny}},
  \bibinfo{journal}{Phys. Rev. Lett.} \textbf{\bibinfo{volume}{98}},
  \bibinfo{pages}{266802} (\bibinfo{year}{2007}).

\bibitem[{\citenamefont{Kuttge et~al.}(2010)\citenamefont{Kuttge, Garc\'ia~de
  Abajo, and Polman}}]{kuttge10}
\bibinfo{author}{\bibfnamefont{M.}~\bibnamefont{Kuttge}},
  \bibinfo{author}{\bibfnamefont{F.~J.} \bibnamefont{Garc\'ia~de Abajo}},
  \bibnamefont{and} \bibinfo{author}{\bibfnamefont{A.}~\bibnamefont{Polman}},
  \bibinfo{journal}{Nano Lett.} \textbf{\bibinfo{volume}{10}},
  \bibinfo{pages}{1537} (\bibinfo{year}{2010}).

\bibitem[{\citenamefont{Volkov and Mikhailov}(1988)}]{volkov88}
\bibinfo{author}{\bibfnamefont{V. A.}~\bibnamefont{Volkov}} \bibnamefont{and}
  \bibinfo{author}{\bibfnamefont{S. A.}~\bibnamefont{Mikhailov}},
  \bibinfo{journal}{Sov. Phys. JETP} \textbf{\bibinfo{volume}{67}},
  \bibinfo{pages}{1639} (\bibinfo{year}{1988}).

\bibitem[{\citenamefont{Wang et~al.}(2011)\citenamefont{Wang, Apell, and
  Kinaret}}]{wang11}
\bibinfo{author}{\bibfnamefont{W.}~\bibnamefont{Wang}},
  \bibinfo{author}{\bibfnamefont{P.}~\bibnamefont{Apell}}, \bibnamefont{and}
  \bibinfo{author}{\bibfnamefont{J.}~\bibnamefont{Kinaret}},
  \bibinfo{journal}{Phys. Rev. B} \textbf{\bibinfo{volume}{84}},
  \bibinfo{pages}{085423} (\bibinfo{year}{2011}).

\bibitem[{\citenamefont{Schmidt et~al.}(2014)\citenamefont{Schmidt, Ditlbacher,
  Hohenester, Hohenau, Hofer, and Krenn}}]{schmidt14}
\bibinfo{author}{\bibfnamefont{F.-P.} \bibnamefont{Schmidt}},
  \bibinfo{author}{\bibfnamefont{H.}~\bibnamefont{Ditlbacher}},
  \bibinfo{author}{\bibfnamefont{U.}~\bibnamefont{Hohenester}},
  \bibinfo{author}{\bibfnamefont{A.}~\bibnamefont{Hohenau}},
  \bibinfo{author}{\bibfnamefont{F.}~\bibnamefont{Hofer}}, \bibnamefont{and}
  \bibinfo{author}{\bibfnamefont{J.~R.} \bibnamefont{Krenn}},
  \bibinfo{journal}{Nat. Commun.} \textbf{\bibinfo{volume}{5}},  
  \bibinfo{pages}{3604}
  (\bibinfo{year}{2014}).

\bibitem[{\citenamefont{Gordon}(2006)}]{gordon06}
\bibinfo{author}{\bibfnamefont{R.}~\bibnamefont{Gordon}},
  \bibinfo{journal}{Phys. Rev. B} \textbf{\bibinfo{volume}{73}},
  \bibinfo{pages}{153405} (\bibinfo{year}{2006}).

\bibitem[{\citenamefont{Barnard et~al.}(2008)\citenamefont{Barnard, White,
  Chandran, and Brongersma}}]{barnard08}
\bibinfo{author}{\bibfnamefont{E.~S.} \bibnamefont{Barnard}},
  \bibinfo{author}{\bibfnamefont{J.~S.} \bibnamefont{White}},
  \bibinfo{author}{\bibfnamefont{A.}~\bibnamefont{Chandran}}, \bibnamefont{and}
  \bibinfo{author}{\bibfnamefont{M.~L.} \bibnamefont{Brongersma}},
  \bibinfo{journal}{Opt. Express} \textbf{\bibinfo{volume}{16}},
  \bibinfo{pages}{16529} (\bibinfo{year}{2008}).

\bibitem[{\citenamefont{Filter et~al.}(2012)\citenamefont{Filter, Qi,
  Rockstuhl, and Lederer}}]{filter12}
\bibinfo{author}{\bibfnamefont{R.}~\bibnamefont{Filter}},
  \bibinfo{author}{\bibfnamefont{J.}~\bibnamefont{Qi}},
  \bibinfo{author}{\bibfnamefont{C.}~\bibnamefont{Rockstuhl}},
  \bibnamefont{and} \bibinfo{author}{\bibfnamefont{F.}~\bibnamefont{Lederer}},
  \bibinfo{journal}{Phys. Rev. B} \textbf{\bibinfo{volume}{85}},
  \bibinfo{pages}{125429} (\bibinfo{year}{2012}).

\bibitem[{\citenamefont{Tserkezis et~al.}(2015)\citenamefont{Tserkezis,
  Esteban, Sigle, Mertens, Herrmann, Baumberg, and Aizpurua}}]{tzerkezis15}
\bibinfo{author}{\bibfnamefont{C.}~\bibnamefont{Tserkezis}},
  \bibinfo{author}{\bibfnamefont{R.}~\bibnamefont{Esteban}},
  \bibinfo{author}{\bibfnamefont{D.~O.} \bibnamefont{Sigle}},
  \bibinfo{author}{\bibfnamefont{J.}~\bibnamefont{Mertens}},
  \bibinfo{author}{\bibfnamefont{L.~O.} \bibnamefont{Herrmann}},
  \bibinfo{author}{\bibfnamefont{J.~J.} \bibnamefont{Baumberg}},
  \bibnamefont{and} \bibinfo{author}{\bibfnamefont{J.}~\bibnamefont{Aizpurua}},
  \bibinfo{journal}{Phys. Rev. A} \textbf{\bibinfo{volume}{92}},
  \bibinfo{pages}{053811} (\bibinfo{year}{2015}).

\bibitem[{\citenamefont{Wang and Shen}(2006)}]{wang06c}
\bibinfo{author}{\bibfnamefont{F.}~\bibnamefont{Wang}} \bibnamefont{and}
  \bibinfo{author}{\bibfnamefont{Y.~R.} \bibnamefont{Shen}},
  \bibinfo{journal}{Phys. Rev. Lett.} \textbf{\bibinfo{volume}{97}},
  \bibinfo{pages}{206806} (\bibinfo{year}{2006}).

\bibitem[{\citenamefont{Novotny and Hecht}(2006)}]{novotnyBook}
\bibinfo{author}{\bibfnamefont{L.}~\bibnamefont{Novotny}} \bibnamefont{and}
  \bibinfo{author}{\bibfnamefont{B.}~\bibnamefont{Hecht}},
  \emph{\bibinfo{title}{\textit{Principles of Nano-Optics}}}
  (\bibinfo{publisher}{Cambridge University Press, Cambridge},
  \bibinfo{year}{2006}).

\bibitem[{\citenamefont{T{\"o}rm{\"a} and Barnes}(2014)}]{torma2014}
\bibinfo{author}{\bibfnamefont{P.}~\bibnamefont{T{\"o}rm{\"a}}}
  \bibnamefont{and} \bibinfo{author}{\bibfnamefont{W.~L.}
  \bibnamefont{Barnes}}, \bibinfo{journal}{Reports on Progress in Physics}
  \textbf{\bibinfo{volume}{78}}, \bibinfo{pages}{013901}
  (\bibinfo{year}{2015}).

\bibitem[{\citenamefont{Politano et~al.}(2015)\citenamefont{Politano, Silkin,
  Nechaev, Vitiello, Viti, Aliev, Babanly, Chiarello, Echenique, and
  Chulkov}}]{politano15}
\bibinfo{author}{\bibfnamefont{A.}~\bibnamefont{Politano}},
  \bibinfo{author}{\bibfnamefont{V.~M.} \bibnamefont{Silkin}},
  \bibinfo{author}{\bibfnamefont{I.~A.} \bibnamefont{Nechaev}},
  \bibinfo{author}{\bibfnamefont{M.~S.} \bibnamefont{Vitiello}},
  \bibinfo{author}{\bibfnamefont{L.}~\bibnamefont{Viti}},
  \bibinfo{author}{\bibfnamefont{Z.~S.} \bibnamefont{Aliev}},
  \bibinfo{author}{\bibfnamefont{M.~B.} \bibnamefont{Babanly}},
  \bibinfo{author}{\bibfnamefont{G.}~\bibnamefont{Chiarello}},
  \bibinfo{author}{\bibfnamefont{P.~M.} \bibnamefont{Echenique}},
  \bibnamefont{and} \bibinfo{author}{\bibfnamefont{E.~V.}
  \bibnamefont{Chulkov}}, \bibinfo{journal}{Phys. Rev. Lett.}
  \textbf{\bibinfo{volume}{115}}, \bibinfo{pages}{216802}
  (\bibinfo{year}{2015}).

\bibitem[{\citenamefont{Thongrattanasiri
  et~al.}(2012)\citenamefont{Thongrattanasiri, Manjavacas, and Garc\'{i}a~de
  Abajo}}]{G.Abajo.ACSNano}
\bibinfo{author}{\bibfnamefont{S.}~\bibnamefont{Thongrattanasiri}},
  \bibinfo{author}{\bibfnamefont{A.}~\bibnamefont{Manjavacas}},
  \bibnamefont{and} \bibinfo{author}{\bibfnamefont{F.~J.}
  \bibnamefont{Garc\'{i}a~de Abajo}}, \bibinfo{journal}{ACS Nano}
  \textbf{\bibinfo{volume}{6}}, \bibinfo{pages}{1766} (\bibinfo{year}{2012}).

\bibitem[{\citenamefont{Virk and Yazyev}(2016)}]{virk2016dirac}
\bibinfo{author}{\bibfnamefont{N.}~\bibnamefont{Virk}} \bibnamefont{and}
  \bibinfo{author}{\bibfnamefont{O.~V.} \bibnamefont{Yazyev}},
  \bibinfo{journal}{Sci. Rep.} \textbf{\bibinfo{volume}{6}},
  \bibinfo{pages}{20220} (\bibinfo{year}{2016}).

\bibitem[{\citenamefont{Novotny et~al.}(1997)\citenamefont{Novotny, Hecht, and
  Pohl}}]{novotny97}
\bibinfo{author}{\bibfnamefont{L.}~\bibnamefont{Novotny}},
  \bibinfo{author}{\bibfnamefont{B.}~\bibnamefont{Hecht}}, \bibnamefont{and}
  \bibinfo{author}{\bibfnamefont{D.~W.} \bibnamefont{Pohl}},
  \bibinfo{journal}{J. Appl. Phys.} \textbf{\bibinfo{volume}{81}},
    \bibinfo{pages}{1798}  (\bibinfo{year}{1997}).

\bibitem[{\citenamefont{Huber et~al.}(2005)\citenamefont{Huber, Ocelic,
  Kazantsev, and Hillenbrand}}]{ocelic05}
\bibinfo{author}{\bibfnamefont{A.}~\bibnamefont{Huber}},
  \bibinfo{author}{\bibfnamefont{N.}~\bibnamefont{Ocelic}},
  \bibinfo{author}{\bibfnamefont{D.}~\bibnamefont{Kazantsev}},
  \bibnamefont{and}
  \bibinfo{author}{\bibfnamefont{R.}~\bibnamefont{Hillenbrand}},
  \bibinfo{journal}{Appl. Phys. Lett.} \textbf{\bibinfo{volume}{87}},
  \bibinfo{eid}{081103} (\bibinfo{year}{2005}).

\bibitem[{\citenamefont{Stauber et~al.}(2018)\citenamefont{Stauber, Low, and
  G\'omez-Santos}}]{Stauber_PRL_2018}
\bibinfo{author}{\bibfnamefont{T.}~\bibnamefont{Stauber}},
  \bibinfo{author}{\bibfnamefont{T.}~\bibnamefont{Low}}, \bibnamefont{and}
  \bibinfo{author}{\bibfnamefont{G.}~\bibnamefont{G\'omez-Santos}},
  \bibinfo{journal}{Phys. Rev. Lett.} \textbf{\bibinfo{volume}{120}},
  \bibinfo{pages}{046801} (\bibinfo{year}{2018}).

\bibitem[{\citenamefont{Bellessa et~al.}(2004)\citenamefont{Bellessa, Bonnand,
  Plenet, and Mugnier}}]{bellessa04}
\bibinfo{author}{\bibfnamefont{J.}~\bibnamefont{Bellessa}},
  \bibinfo{author}{\bibfnamefont{C.}~\bibnamefont{Bonnand}},
  \bibinfo{author}{\bibfnamefont{J.~C.} \bibnamefont{Plenet}},
  \bibnamefont{and} \bibinfo{author}{\bibfnamefont{J.}~\bibnamefont{Mugnier}},
  \bibinfo{journal}{Phys. Rev. Lett.} \textbf{\bibinfo{volume}{93}},
  \bibinfo{pages}{036404} (\bibinfo{year}{2004}).

\bibitem[{\citenamefont{Tr\"ugler and Hohenester}(2008)}]{trugler08}
\bibinfo{author}{\bibfnamefont{A.}~\bibnamefont{Tr\"ugler}} \bibnamefont{and}
  \bibinfo{author}{\bibfnamefont{U.}~\bibnamefont{Hohenester}},
  \bibinfo{journal}{Phys. Rev. B} \textbf{\bibinfo{volume}{77}},
  \bibinfo{pages}{115403} (\bibinfo{year}{2008}).

\bibitem[{\citenamefont{Schmidt et~al.}(2016)\citenamefont{Schmidt, Esteban,
  Gonz\'alez-Tudela, Giedke, and Aizpurua}}]{schmidt15}
\bibinfo{author}{\bibfnamefont{M.~K.} \bibnamefont{Schmidt}},
  \bibinfo{author}{\bibfnamefont{R.}~\bibnamefont{Esteban}},
  \bibinfo{author}{\bibfnamefont{A.}~\bibnamefont{Gonz\'alez-Tudela}},
  \bibinfo{author}{\bibfnamefont{G.}~\bibnamefont{Giedke}}, \bibnamefont{and}
  \bibinfo{author}{\bibfnamefont{J.}~\bibnamefont{Aizpurua}},
  \bibinfo{journal}{ACS Nano} \textbf{\bibinfo{volume}{10}},
  \bibinfo{pages}{6291} (\bibinfo{year}{2016}).

\bibitem[{\citenamefont{Mermin}(1970)}]{Mermin.PRB}
\bibinfo{author}{\bibfnamefont{N.~D.} \bibnamefont{Mermin}},
  \bibinfo{journal}{Phys. Rev. B} \textbf{\bibinfo{volume}{1}},
  \bibinfo{pages}{2362} (\bibinfo{year}{1970}).

\bibitem[{\citenamefont{Wunsch et~al.}(2006)\citenamefont{Wunsch, Stauber,
  Sols, and Guinea}}]{Wunsch.NJP}
\bibinfo{author}{\bibfnamefont{B.}~\bibnamefont{Wunsch}},
  \bibinfo{author}{\bibfnamefont{T.}~\bibnamefont{Stauber}},
  \bibinfo{author}{\bibfnamefont{F.}~\bibnamefont{Sols}}, \bibnamefont{and}
  \bibinfo{author}{\bibfnamefont{F.}~\bibnamefont{Guinea}},
  \bibinfo{journal}{New J. Phys.} \textbf{\bibinfo{volume}{8}},
  \bibinfo{pages}{318} (\bibinfo{year}{2006}).

\bibitem[{\citenamefont{Hwang and Das~Sarma}(2007)}]{Hwang_PRB_2007}
\bibinfo{author}{\bibfnamefont{E.~H.} \bibnamefont{Hwang}} \bibnamefont{and}
  \bibinfo{author}{\bibfnamefont{S.}~\bibnamefont{Das~Sarma}},
  \bibinfo{journal}{Phys. Rev. B} \textbf{\bibinfo{volume}{75}},
  \bibinfo{pages}{205418} (\bibinfo{year}{2007}).

\bibitem[{\citenamefont{Eremeev et~al.}(2015)\citenamefont{Eremeev, Tsirkin,
  Nechaev, Echenique, and Chulkov}}]{eremeev15}
\bibinfo{author}{\bibfnamefont{S.~V.} \bibnamefont{Eremeev}},
  \bibinfo{author}{\bibfnamefont{S.~S.} \bibnamefont{Tsirkin}},
  \bibinfo{author}{\bibfnamefont{I.~A.} \bibnamefont{Nechaev}},
  \bibinfo{author}{\bibfnamefont{P.~M.} \bibnamefont{Echenique}},
  \bibnamefont{and} \bibinfo{author}{\bibfnamefont{E.~V.}
  \bibnamefont{Chulkov}}, \bibinfo{journal}{Sci. Rep.}
  \textbf{\bibinfo{volume}{5}}, \bibinfo{pages}{12819} (\bibinfo{year}{2015}).

\bibitem[{\citenamefont{Scharf and Matos-Abiague}(2012)}]{Scharf_PRB_2012}
\bibinfo{author}{\bibfnamefont{B.}~\bibnamefont{Scharf}} \bibnamefont{and}
  \bibinfo{author}{\bibfnamefont{A.}~\bibnamefont{Matos-Abiague}},
  \bibinfo{journal}{Phys. Rev. B} \textbf{\bibinfo{volume}{86}},
  \bibinfo{pages}{115425} (\bibinfo{year}{2012}).

\bibitem[{\citenamefont{Principi et~al.}(2012)\citenamefont{Principi, Carrega,
  Asgari, Pellegrini, and Polini}}]{Principi_PRB_2012}
\bibinfo{author}{\bibfnamefont{A.}~\bibnamefont{Principi}},
  \bibinfo{author}{\bibfnamefont{M.}~\bibnamefont{Carrega}},
  \bibinfo{author}{\bibfnamefont{R.}~\bibnamefont{Asgari}},
  \bibinfo{author}{\bibfnamefont{V.}~\bibnamefont{Pellegrini}},
  \bibnamefont{and} \bibinfo{author}{\bibfnamefont{M.}~\bibnamefont{Polini}},
  \bibinfo{journal}{Phys. Rev. B} \textbf{\bibinfo{volume}{86}},
  \bibinfo{pages}{085421} (\bibinfo{year}{2012}).

\bibitem[{\citenamefont{Qi et~al.}(2014)\citenamefont{Qi, Liu, and
  Xie}}]{QiPRB}
\bibinfo{author}{\bibfnamefont{J.}~\bibnamefont{Qi}},
  \bibinfo{author}{\bibfnamefont{H.}~\bibnamefont{Liu}}, \bibnamefont{and}
  \bibinfo{author}{\bibfnamefont{X.~C.} \bibnamefont{Xie}},
  \bibinfo{journal}{Phys. Rev. B} \textbf{\bibinfo{volume}{89}},
  \bibinfo{pages}{155420} (\bibinfo{year}{2014}).

\bibitem[{\citenamefont{Aizpurua et~al.}(2008)\citenamefont{Aizpurua, Taubner,
  Garc\'{i}a~de Abajo, Brehm, and Hillenbrand}}]{Aiz_sub_enh}
\bibinfo{author}{\bibfnamefont{J.}~\bibnamefont{Aizpurua}},
  \bibinfo{author}{\bibfnamefont{T.}~\bibnamefont{Taubner}},
  \bibinfo{author}{\bibfnamefont{F.~J.} \bibnamefont{Garc\'{i}a~de Abajo}},
  \bibinfo{author}{\bibfnamefont{M.}~\bibnamefont{Brehm}}, \bibnamefont{and}
  \bibinfo{author}{\bibfnamefont{H.}~\bibnamefont{Hillenbrand}},
  \bibinfo{journal}{Opt. Express} \textbf{\bibinfo{volume}{16}},
  \bibinfo{pages}{1529} (\bibinfo{year}{2008}).

\bibitem[{\citenamefont{Villanova and Park}(2016)}]{PhysRevB.93.085122}
\bibinfo{author}{\bibfnamefont{J.~W.} \bibnamefont{Villanova}}
  \bibnamefont{and} \bibinfo{author}{\bibfnamefont{K.}~\bibnamefont{Park}},
  \bibinfo{journal}{Phys. Rev. B} \textbf{\bibinfo{volume}{93}},
  \bibinfo{pages}{085122} (\bibinfo{year}{2016}).

\bibitem[{\citenamefont{Moon et~al.}(2011)\citenamefont{Moon, Han, Lee, and
  Choi}}]{PhysRevB.84.195425}
\bibinfo{author}{\bibfnamefont{C.-Y.} \bibnamefont{Moon}},
  \bibinfo{author}{\bibfnamefont{J.}~\bibnamefont{Han}},
  \bibinfo{author}{\bibfnamefont{H.}~\bibnamefont{Lee}}, \bibnamefont{and}
  \bibinfo{author}{\bibfnamefont{H.~J.} \bibnamefont{Choi}},
  \bibinfo{journal}{Phys. Rev. B} \textbf{\bibinfo{volume}{84}},
  \bibinfo{pages}{195425} (\bibinfo{year}{2011}).

\end{thebibliography}

\end{document}